\documentclass[fleqn,usenatbib]{mnras}

\usepackage{newtxtext,newtxmath}
\usepackage[T1]{fontenc}
\usepackage{xcolor}
\usepackage[normalem]{ulem}

\DeclareRobustCommand{\VAN}[3]{#2}
\let\VANthebibliography\thebibliography
\def\thebibliography{\DeclareRobustCommand{\VAN}[3]{##3}\VANthebibliography}

\usepackage{graphicx}	% Including figure files
\usepackage{amsmath}	% Advanced maths commands
\usepackage{subfigure}  % For \subfigure

\newcommand{\Msun}[1]{$\rm{M}_{\odot}$#1}
\definecolor{darkgreen}{rgb}{0.0, 0.5, 0.0} % custom dark green

\title[]{The impact of rotational mixing in intermediate-age star clusters with extended main-sequence turn-offs and extended red clumps}

\author[L. Martinelli et al.]{L. Martinelli,$^{1}$\thanks{E-mail: lorenzo.martinelli@uon.edu.au}
A. Miglio$^{2, 3}$,
G. Buldgen$^{4}$,
H. Schunker$^{1}$,
C. Georgy$^{6}$,
G. Cordoni$^{7}$,
K. Brogaard$^{5, 2}$,
\newauthor 
P. Eggenberger$^{6}$,
E. Farrell$^{6}$
\\
$^{1}$ The University of Newcastle, University Dr, Callaghan NSW 2308, Australia\\
$^{2}$Dipartimento di Fisica e Astronomia, Università degli Studi di Bologna, Via Gobetti 93/2, I-40129 Bologna, Italy\\
$^{3}$INAF - Osservatorio di Astrofisica e Scienza dello Spazio di Bologna, Via Gobetti 93/3, I-40129 Bologna, Italy\\
$^{4}$STAR Institute, Université de Liège, Liège, Belgium\\
$^{5}$Stellar Astrophysics Centre, Department of Physics \& Astronomy, Aarhus University, Ny Munkegade 120, 8000 Aarhus C, Denmark\\
$^{6}$Département d’astronomie, Université de Genève, Chemin Pegasi 51, CH-1290 Versoix, Switzerland\\
$^{7}$Research School of Astronomy and Astrophysics, The Australian National University, Canberra, ACT 2611, Australia\\
}

\date{Accepted XXX. Received YYY; in original form ZZZ}

\pubyear{2025}

\usepackage[mathlines]{lineno}
\begin{document}
\label{firstpage}
\pagerange{\pageref{firstpage}--\pageref{lastpage}}
\maketitle

\begin{abstract}
The extended main-sequence turn-offs (eMSTOs) and extended red clumps (eRCs) observed in intermediate-age star clusters challenge the traditional understanding of clusters as simple stellar populations. Recently, eMSTOs have been interpreted as signatures of stellar rotation. In this work, we test the effectiveness of rotational mixing in shaping the colour-magnitude diagram (CMD) of star clusters.
We computed a set of separate single-age synthetic stellar populations, referred to as "Base Stellar Populations" (BSPs), including stellar rotation. These BSPs were generated from two grids of stellar models that share the same input physics but differ in the efficiency of rotational mixing. We used an optimization algorithm to determine the best combination of BSPs to fit the CMDs of two star clusters: the Small Magellanic Cloud cluster NGC~419 and the Milky Way cluster NGC~1817.
The synthetic clusters with weak rotational mixing provide the best fit to both the eMSTO and eRC features for both clusters, and are consistent with the luminosities and asteroseismic masses we derived for eRC stars in NGC~1817. In contrast, synthetic clusters with strong rotational mixing result in overly bright post-main-sequence stars, inconsistent with observations. This suggests that, for intermediate-mass stars, the influence of rotational mixing of chemical elements on stellar evolution cannot be so strong as to significantly increase the post–main-sequence luminosity. A simple test suggests that accounting for self-extinction by decretion discs in equator-on fast rotators could influence inferred rotation distributions and help reconcile the projected rotational velocity discrepancy across the eMSTO between models and observations.
\end{abstract}


%%%%%%%%%%%%%%%%%%%%%%%%%%%%%%%%%%%%%%%%%%%%%%%%%%

%%%%%%%%%%%%%%%%% BODY OF PAPER %%%%%%%%%%%%%%%%%%
\section{Introduction}\label{introduction}
In the last two decades, high-precision \textit{Hubble Space Telescope} (HST) photometry of star clusters has revealed that the colour-magnitude digram (CMD) of most clusters in the Magellanic Clouds younger than \textasciitilde2.5~Gyr exhibit extended main sequence turn-offs (eMSTOs)  \citep[][]{Milone+2009, Rubele+2010, Girardi+2013, Milone+2023}, with some also displaying extended (or dual) red clumps (eRCs) \citep[][]{Girardi+2009, Milone+2009}. More recently, high-precision astrometric and photometric data from the \emph{Gaia} mission \citep[][]{GaiaEDR32021, GaiaDR22018, GaiaDR32023} have shown that these features are also present in many Galactic open clusters \citep[][]{Cordoni+2018, Sun+2019, Cordoni+2024}. 
These features are observed in clusters of both the Magellanic Clouds and the Milky Way, suggesting a shared underlying physical cause \citep[][]{Li+2024}.

The origin of the eMSTO was initially attributed to an age spread (on the order of hundreds of Myr) within the stellar populations of these clusters. This scenario also accounts for the presence of eRCs in some clusters \citep[][]{Rubele+2010, Girardi+2013}, since a spread in age  corresponds to a spread of mass among the stars in the same evolutionary phase. 

However, the age spread scenario directly challenges the conventional understanding of star formation. Initial stellar feedback is expected to rapidly disperse  surrounding gas  \citep[e.g.,][]{Krumholz2014}, and star formation in a typical \textasciitilde$10^5$ M$_{\odot}$ star cluster should conclude in $\lesssim 10^7$ years \citep[][]{Elmegreen&Efremov1997} and so an age spread of hundreds of Myr is not expected.
Although scenarios involving mergers between star clusters with different ages, or between a cluster and a giant molecular cloud have been proposed to account for age spreads \citep[][]{Bekki+2009}, they are expected to be rare and thus fail to explain the pervasiveness of the eMSTO feature. Furthermore, the age spread scenario cannot explain other CMD features that are consistent with a single stellar population  \citep[e.g.][]{Li+2014a, Li+2014b,Bastian&Niederhofer2015, Li+2016b, Wu+2016}. 
For instance, while the model explains some clusters with eRCs, it struggles with clusters with compact RCs \citep[][]{Li+2014b}. Finally, clusters with eMSTOs often have a narrow sub-giant branch (SGB), inconsistent with a large internal age spread \citep[][]{Li+2014a, Li+2016b, Bastian&Niederhofer2015, Wu+2016}.

High-resolution spectroscopy of clusters with an eMSTO and eRC have revealed a systematic difference in the projected rotation velocities  ($v \sin i$, where $v$ is the surface velocity and $i$ is the stellar inclination) between stars in the red and blue regions of the eMSTO \citep[][]{Marino+2018a, Marino+2018b, Kamann+2018, Bastian+2018, Sun+2019, Kamann+2020}. This suggests that stellar rotation is a plausible cause of the effects leading to eMSTOs.

According to the stellar rotation scenario, the broadening of the MSTO is primarily driven by \emph{gravity darkening} \citep[][]{VonZeipel1924, Espinosa-Lara&Rieutord2011}, where the equator of a rapidly rotating star appears darker and cooler than its poles due to the variation in surface gravity caused by the centrifugal acceleration across the star’s surface. This effect has been directly observed in some B- and F-type stars using interferometric imaging \citep[][]{vanBelle2012}. Consequently, the measured effective temperature, and thus the color and magnitude of a star, depend on the rotation rate and the inclination angle of the rotation axis relative to the line-of-sight.

\citet{Bastian&deMink2009} were the first to demonstrate that rotation could, at least partially, account for the eMSTOs.
Rotation causes hydrodynamical instabilities and large-scale plasma motions in radiative regions, which is thought to induce the mixing of stellar material and transport of angular momentum and chemical elements \citep[][]{Maeder&Meynet2000, Maeder2009}. However, \citet{Girardi+2011} argued that rotational mixing and gravity darkening exert opposing influences on a cluster’s CMD, rendering their combined effects insufficient to explain the eMSTO width. However, their analysis was constrained to rotational speeds not exceeding 70~\% of the critical velocity and neglected the effect of the inclination angle.
Subsequent investigations, incorporating stars with rotation rates approaching critical velocities and the effect of inclination angles \citep[e.g.,][]{Brandt&Huang2015, Cordoni+2018, Gossage+2019, Dresbach+2023, Cordoni+2024}, have shown that in some cases a distribution of rotation rates alone can qualitatively reproduce the observed eMSTOs. 
Additionally, the stellar rotation scenario can explain why clusters with eMSTOs exhibit a narrow SGB \citep[][]{Li+2014a, Bastian&Niederhofer2015, Li+2016a, Wu+2016}: as the rotation rates of subgiant stars would rapidly decrease, leading to a morphology of the CMD feature that remains consistent with a single stellar population, provided that rotation-induced mixing does not significantly affect the lifetime and luminosity distribution of rotating stars (see Section \ref{subsubsec:rot_mix}). 
This observation may offer an opportunity to constrain the extent of rotation-induced mixing in stars, which remains one of the major sources of uncertainty in modelling the effects of stellar rotation \citep[][]{Salaris&Cassisi2017}.
In clusters hosting B-type main-sequence stars, a non-negligible fraction exhibit Balmer emission. These stars, referred to as Be stars, are thought to be surrounded by optically thin circumstellar discs formed by rapid stellar rotation, with the emission lines originating from the disc \citep[][]{Slettebak1988,Porter&Rivinius2003,McSwain+2005,Martayan+2008,Rivinius+2013, Bastian+2016, Milone+2018}. The fraction of Be stars varies with spectral type. Recent studies have also revealed a population of UV-dim stars in older clusters such as NGC 1783 \citep{Martocchia+2023}, which show reduced UV flux compared to normal A/F-type stars. In younger clusters, many Be stars display a similar UV-dim characteristic, suggesting a connection between circumstellar discs and UV attenuation \citep[][]{Milone+2023b, Li+2024}. Fast rotators may develop decretion discs in which dust condenses in the outer regions, and in clusters older than approximately 1 Gyr the stars are not hot enough to ionize the disc material, so the discs may be detectable only through absorption signatures \citep[][]{D'Antona+2023, Kamann+2023, He+2025}.

In 1-D stellar evolution codes stellar rotation is treated according to the shellular rotation approximation \citep[][]{Kippenhahn&Thomas1970, Endal&Sofia1976, Meynet&Maeder1997}. Within the shellular approximation the transport of angular momentum is treated in two different ways: using a simple diffusion equation \citep[][]{Endal&Sofia1978, Pinsonneault+1989, Heger+2000} such as in codes like \verb|MESA| \citep[][]{Paxton+2011, Paxton+2013, Paxton+2015, Paxton+2018, Paxton+2019, Jermyn+2023} and \verb|PARSEC| \citep[][]{Nguyen+2022}, KEPLER \citep[][]{Weaver+1978}, STERN \citep[][]{Heger+2000}, or using an advecto-diffusive equation \citep[][]{Zahn1992, Meynet&Maeder1997, Maeder&Zahn1998, Maeder2009} that consider the advective nature of the meridional flows adopted by \verb|GENEC| \citep[][]{Eggenberger+2008}, \verb|STAREVOL| \citep[][]{Palacios+2006, Decressin+2009}, \verb|CESTAM| \citep[][]{Marques+2013}, and \verb|FRANEC| \citep[][]{Chieffi&Limongi2013}. Different stellar evolutionary models can also differ in the derivation of the diffusion coefficients for the rotationally induced instabilities, often leading to significantly different efficiencies of these processes under the same initial stellar parameters \citep[][]{Nandal+2024}. \verb|MESA| follows the formulation of \citet{Heger+2000}, using a highly simplified order-of-magnitude approach, whereas \verb|GENEC| and \verb|PARSEC V2.0| adopt a more physical description that prioritize self-consistency \citet{Chaboyer&Zahn1992, Talon&Zahn1997, Meynet&Maeder1997, Maeder2009}. 

Additionally, current rotating models fail to predict the evolution of the internal rotation profile as inferred from seismic constraints, indicating that additional angular momentum transport processes are required beyond those currently implemented in standard stellar evolution codes \citep[e.g.][]{Eggenberger+2012b, Beck+2012,Eggenberger+2019,Aerts+2019}. This demonstrates that current rotating models are incomplete in their treatment of angular momentum transport, and highlights the need to constrain rotational mixing, whose effects are harder to isolate due to degeneracies with convective overshooting, composition, and age \citep[][]{Eggenberger+2011}.

The goal of this work is to test how different formulations of rotation-induced mixing, which predict significantly different mixing efficiencies, affect the CMD of intermediate-age synthetic star clusters with a distribution of initial stellar rotation rates. Using the \verb|MESA| stellar evolution code, we compute two grids of stellar models adopting, respectively, the default diffusive formulation and an approximation of the advective-diffusive formulation as implemented in \verb|GENEC| models from \citet{Georgy+2013}. 

This paper is organized as follows. In Section~\ref{sec:models_grids}, we describe the input physics of the stellar models, including the implementation of rotation and the prescriptions for rotation-induced mixing adopted in our grids. Section~\ref{sec:iso_clust_construction} outlines the method used to find the best-fit synthetic clusters of observed star clusters. In Section~\ref{sec:rot_mix_eMSTOs_eRCs}, we examine the impact of rotational mixing on the CMD features of synthetic clusters and analyse and discuss the properties of the best-fit models for NGC~419 and NGC~1817. Our conclusions are presented in Section~\ref{sec:conclusions}.
% End of Introduction

\section{Grids of Stellar Models with Rotation}\label{sec:models_grids}
In this section, we describe the physics included in the stellar evolutionary models and the implementation of stellar rotation.  
We use the publicly available, open-source, one-dimensional stellar evolution code, \emph{Modules for Experiments in Stellar Astrophysics} \citep[\texttt{MESA}][]{Paxton+2011, Paxton+2013, Paxton+2015, Paxton+2018, Paxton+2019, Jermyn+2023}, version \verb'r-24.03.1'.  
All models start from the pre-main sequence and extend up to the first few pulses of the thermally pulsating asymptotic giant branch.
We compute two grids of models, that differ only in the implementation of rotation-induced mixing of chemical elements and convective core overshooting, referred to as \verb|grid 1| and \verb|grid 2|, and described in detail in Sections~\ref{subsubsec:conv_ovsh} and \ref{subsubsec:rot_mix}.  
The grids of stellar models span the mass range 0.6~–~4.0~\Msun{}, with a step of 0.1~\Msun{} for $0.6 \leq M < 1.7$~\Msun{} and $3.0 < M \leq 4.0$~\Msun{}, and a finer step of 0.025~\Msun{} for models with  $1.7 \leq M \leq 3.0$~\Msun{}. The finer mass resolution between 1.7~–~3.0~\Msun{} is crucial to resolve the transition from degenerate to non-degenerate core helium ignition, which is key to modelling the secondary clump and understanding the eRCs in intermediate-age clusters \citep{Girardi+1998, Girardi1999, Girardi+2000}.

\subsection{Stellar Evolutionary Models}\label{subsec:input_physics}
Here, we  describe the different physics adopted in the stellar models of both \verb|grid 1| and \verb|grid 2|.
The most relevant settings adopted in the stellar evolutionary models are summarized in Table \ref{tab:models_main_settings}, while  the \verb|inlist_project| and \verb|run_star_extras.f90| files containing the complete settings are available on Zenodo: \url{https://doi.org/10.5281/zenodo.15601753}.

\begin{table*}
 \caption{Summary of the adopted physics and characteristics of the stellar models in \texttt{grid 1} and \texttt{grid 2}.}
 \label{tab:models_main_settings}
 \begin{tabular*}{\textwidth}{l@{\hspace*{35pt}}l@{\hspace*{35pt}}l@{\hspace*{35pt}}l@{\hspace*{35pt}}}
  \hline
  Ingredient&Adopted Prescriptions&Reference&Grid\\
  &and Parameters&&\\
  \hline
     Inital chemical composition &$X_{\rm{ini}}=0.7244$, $Y_{\rm{ini}}=0.2611$,&Section \ref{subsubsec:solar_scaled_abundances}&\texttt{1 \& 2}\\
     (solar scaled)&$Z_{\rm{ini}}=0.0145$&\\[3pt]
    Equation of State&default&\citet{Jermyn+2023}\\[3pt]
    Opacity&OPAL Type I for $\log T \geq 3.88$;&\citet{Iglesias+1993, Iglesias+1996}&\texttt{1 \& 2}\\
           & Ferguson for $\log T < 3.88$;&\citet{Ferguson+2005}&\texttt{1 \& 2}\\
           &Type I $\rightarrow$ Type II&\\
           &at the end of H burning&   \\[3pt]
    Nuclear Reaction Network&\texttt{pp\textunderscore and\textunderscore cno\textunderscore extras.net}&&\texttt{1 \& 2}\\
                          &tracks 25 species: &&\\
                          &$^1\rm{H}$, $^2\rm{H}$, $^3\rm{He}$, $^4\rm{He}$, $^7\rm{Li}$, $^7\rm{Be}$, $^8\rm{B}$,&\\
                          & $^{12}\rm{C}$, $^{13}\rm{C}$, $^{13}\rm{N}$, $^{14}\rm{N}$,$^{15}\rm{N}$, $^{14}\rm{O}$,  &\\
                          &$^{15}\rm{O}$, $^{16}\rm{O}$, $^{17}\rm{O}$, $^{18}\rm{O}$, $^{17}\rm{F}$, $^{18}\rm{F}$,  &\\
                          &$^{19}\rm{F}$, $^{18}\rm{Ne}$, $^{19}\rm{Ne}$, $^{20}\rm{Ne}$, $^{22}\rm{Mg}$, $^{24}\rm{Mg}$&\\[3pt]
    Nuclear Reaction Rates&default except for:&\citet{Jermyn+2023}&\texttt{1 \& 2}\\
                          &$^{14}\text{N}(\text{p}, \gamma)^{15}\text{O}$;&\citet{Imbriani+2005}\\
                          &$^{12}\text{C}(\alpha, \gamma)^{16}\text{O}$;&\citet{Kunz+2002}\\[3pt]
    Boundary Conditions&\texttt{Trampedach\textunderscore solar}&\citet{Trampedach+2014, Ball2021}&\texttt{1 \& 2}\\[3pt]
    Element Diffusion&default settings; turned off for&\citet{Jermyn+2023}&\texttt{1 \& 2}\\
                    &vanishing convective envelopes,&\\
                    &stable convective cores and post MS&\\[3pt]
    Rotation&0.0, 0.1, 0.3, 0.5, 0.6, 0.7, 0.8, 0.9, 0.95 $\omega_{\rm{i}}$&Section \ref{subsubsec:rotation}&\texttt{1 \& 2}\\[3pt]
    Convection&\texttt{mlt\textunderscore option = 'ML1'}, $\alpha_{\rm{MLT}}=1.95$&\citet{Bohm-Vitense1958}&\texttt{1 \& 2}\\[3pt]
    Core overshoot&exponential overshooting: $f = 0.0174$ &\citet{Herwig2000,Paxton+2011}&\texttt{1}\\[3pt]
    &Step overshooting: $\alpha_{\rm{ov}}=0.1$&\citet{Paxton+2011, Georgy+2013}&\texttt{2}\\[3pt]
    Envelope undershoot&step undershooting: $\alpha_{\rm{ov}}=0.2$&\citet{Paxton+2011, Khan+2018}&\texttt{1}\\[3pt]
    Rotational Mixing&diffusive implementation: &\citet{Heger+2000};&\texttt{1}\\
    &$f_{c} = 1/30$, $f_{\mu} = 0.05$&\citet{Pinsonneault+1989}&\\[3pt]
    &custom&Appendix \ref{app:rot_mix_calib}&\texttt{2}\\[3pt]
    Mass Loss&MS with rotational boost:&&\\
    &$\eta_{\rm{Dutch}}=1.0$, $\xi=0.43$,&\citet{Paxton+2013,Glebeek+2009}&\\
    &boost factor capped at $10^4$,&\citet{deJager+1988, Langer1998}&\\
    &implicit mass loss to keep $\Omega<\Omega_{\rm{crit}}$&&\texttt{1 \& 2}\\[3pt]
    &RGB: none&&\\[3pt]
    &AGB: $\eta_{\rm{Bloecker}}=0.2$&\citet{Bloecker1995}&\\
  \hline
 \end{tabular*}
\end{table*}

\subsubsection{Solar-scaled abundances}\label{subsubsec:solar_scaled_abundances}
The reference adopted for calculating the solar-scaled abundances is the solar chemical composition from \citet{Asplund+2009}. 
In this work, we adopted \(\Delta Y/\Delta Z = 1.5\), a value consistent with that derived from the protosolar abundances of \citet{Asplund+2009} and within the commonly accepted range \citep[from about 1.4 to 2.5, see][]{Serenelli&Basu}. A similar choice is made by \citet{Choi+2016} and \citet{Ziółkowska+2024}.

We computed models for two different values of the metal mass fraction, \(Z = 0.01\) and \(Z = 0.004\), which approximately match the metallicities of the star clusters NGC~1817 and NGC~419, respectively (see Sections \ref{subsec:NGC419} and \ref{subsec:NGC1817}).

\subsubsection{Microphysics}\label{subsubsec:microphysics} 
We used the default \verb|MESA| equation of state described in \citet{Jermyn+2023}, the OPAL opacity tables \citep{Iglesias&Rogers1993, Iglesias&Rogers1996} for $T \ga 10^4$, and opacity tables from \citet{Ferguson+2005} for $T \la 10^4$.

We selected a nuclear reaction network named \verb|pp_and_cno_extras.net|, that incorporates reactions from the pp~chain and CNO~cycle, as well as those associated with helium burning and the burning of heavier elements.
We adopted nuclear reaction rates relevant for $T < 10^7$~K from the NACRE compilation \citep[][]{Angulo+1999} across all temperatures. For the remaining reactions, we use rates from JINA REACLIB \citep[][]{Cyburt+2010}, or those from \citet{Caughlan&Fowler1988} when not available in either NACRE or JINA REACLIB. Additionally, we updated the specific reaction rates for $^{14}\rm{N}(p,\gamma)^{15}\rm{O}$ from \citet[][]{Imbriani+2005} and $^{12}\rm{C}(\alpha,\gamma)^{16}\rm{O}$ from \citet[][]{Kunz+2002}.

We included atomic diffusion, neglecting radiative acceleration, and left most of the MESA controls relevant to atomic diffusion fixed to their default values. The implementation of atomic diffusion in MESA is described in \citet{Paxton+2015,Paxton+2018}. 
Since we neglected radiative acceleration, atomic diffusion was switched off, to avoid excessive helium and metal depletion in the outer layers, when either the mass fraction of the outer convective envelope fell below 0.5\% of the total stellar mass or the star developed a persistent convective core, consistent with the \texttt{PARSEC} models \citep[][]{Bressan+2012}.

\subsubsection{Model atmosphere}\label{subsubsec:atm}
To model the transition from the opaque depths to the observable outer layers and set the appropriate boundary conditions, we used the \( T(\tau) \) relation from \citet{Ball2021}. This relation is an analytic fit to the radiation-coupled hydrodynamics simulations of near-surface convection with solar parameters by \citet{Trampedach+2014} and is broadly representative of the full grid of simulations from the same work. This choice is implemented in \verb|MESA| using the setting \verb|atm_T_tau_relation = "Trampedach_solar"|.

\subsubsection{Convection and overshooting}\label{subsubsec:conv_ovsh}
Convective mixing in \verb|MESA| is treated as a time-dependent diffusive process with a diffusion coefficient computed within the mixing length theory (MLT) formalism \citep{Bohm-Vitense1958}. We fixed the  free parameter of the MLT, $\alpha_{\rm{MLT}}=1.95$, the initial metallicity $Z_{\rm{ini}}=0.0145$, and the helium mass fraction $Y_{\rm{ini}}=0.2611$ by calibrating a solar model.
We determined the radial boundaries of convective regions using the Schwarzschild criterion except for the convective core boundary during core helium burning, which we treated according to the MOV scheme described in Section~2 of \citet{Bossini+2017}.

In \verb|grid 1|, we used the exponential function for core overshooting prescribed in \citet{Herwig2000} and \citet{Paxton+2011}, using $f_{\rm{ov}} = 0.0174$ \footnote{This value corresponds to $\alpha_{\rm{ov}} \approx 0.2$ in the step overshooting prescription \citep{Claret&Torres2017}.} for stars with $M > 1.1$ \Msun{}. 
Below this mass limit, we did not apply any overshooting following the results of \citet{Claret&Torres2017}. However, we chose not to implement the gradual increase of the overshooting free parameter from 1.1 to 2~\Msun{}, as suggested in the same study, in order to avoid imposing a predefined trend on core boundary mixing efficiency that could obscure the effects of rotation-induced mixing, and to remain consistent with the treatment adopted in \verb|grid 2|.

In \verb|grid 2| we used the step overshooting prescription \citep[][]{Paxton+2011}. For stars with $M>1.1$~\Msun{} we adopted an $\alpha_{\rm{ov}}=0.1$ to be consistent with the choice in the grid of \verb|GENEC| models in \citet{Georgy+2013}, which served as our reference grid for calibrating the efficiency of rotation-induced mixing.  As in \verb|grid 1|, we did not include overshooting for stars $M \leq 1.1$~\Msun{}. This lower amount of overshooting in \verb|grid 2| is compensated by the significantly larger efficiency of rotational mixing in the convective boundary region as in the rotating models of \citet{Georgy+2013}.

In both \verb|grid 1| and \verb|grid 2|, we set the convective core overshoot during the core helium burning phase with a step overshooting with $\alpha_{\rm{ov}}=0.5$ as in \citet{Bossini+2017}.
Below the convective envelope, we implemented an exponential overshooting with $f=0.02$ following \citet{Khan+2018}.

\subsubsection{Rotation}\label{subsubsec:rotation}
\begin{figure*}
    \includegraphics[width=0.8\textwidth]{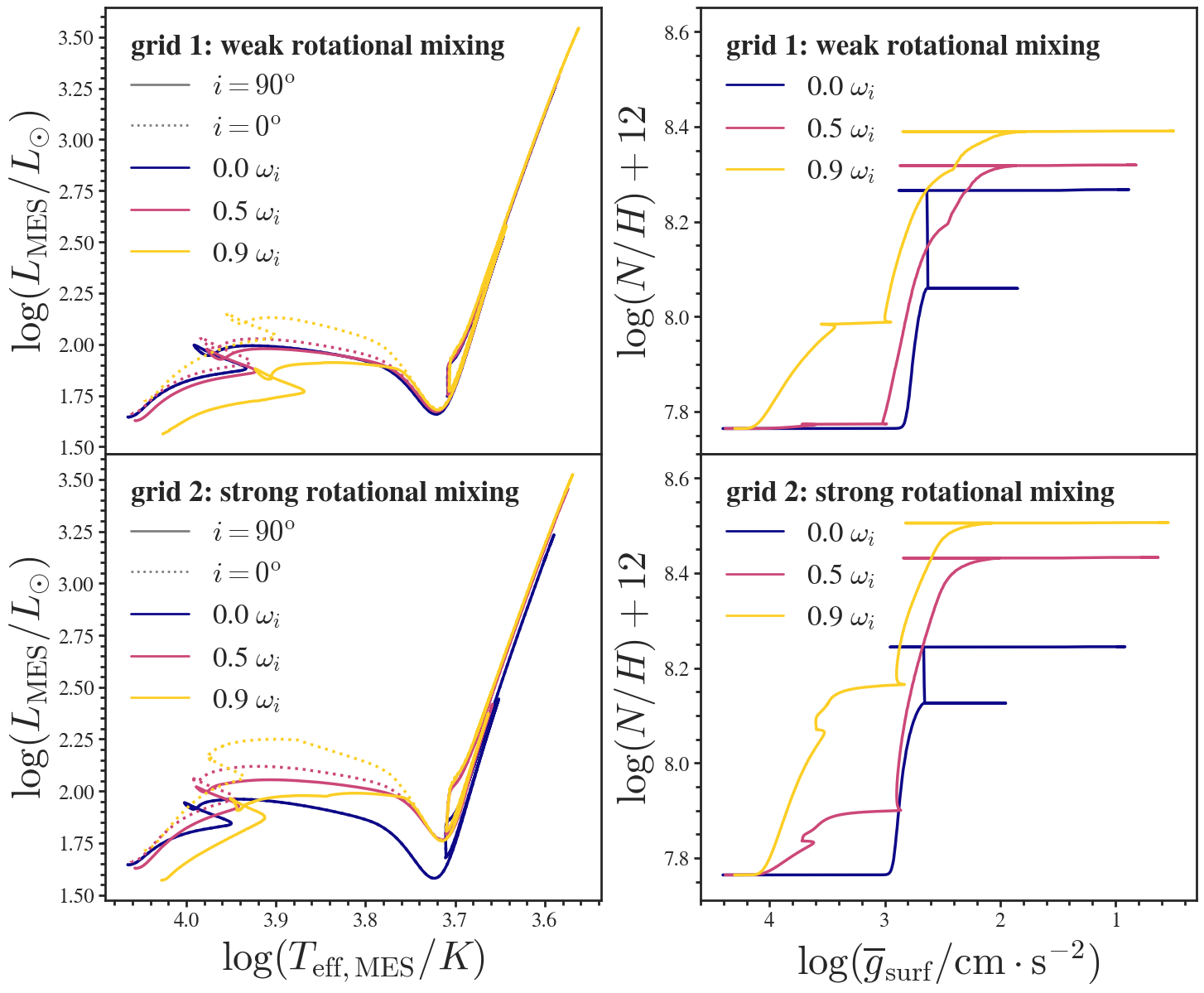}
    \caption{Left: evolutionary tracks of 2.5 \Msun{} models at $Z=0.01$ with initial rotation rates 0.0, 0.5, and 0.9 $\omega_{\rm{i}}$, viewed at an inclination angle $i=90^{\rm{o}}$ (solid) and $i=0^{\rm{o}}$ (dotted), for \texttt{grid 1} (top) and \texttt{grid 2} (bottom). The evolutionary track without rotation in \texttt{grid 1} exhibit a more extended main sequence than that in \texttt{grid 2}, due to the larger convective core overshooting adopted in \texttt{grid 1} (see Section \ref{subsubsec:conv_ovsh}). Gravity darkening on the evolutionary tracks presented in this figure is calculated in a manner consistent with \citet{Espinosa-Lara&Rieutord2011}. Right: evolution of the surface nitrogen-to-hydrogen ratio as a function of the average surface gravity for the same models with \texttt{grid 1} models in the top panel and \texttt{grid 2} models in the bottom panel.}
    \label{fig:grid1_vs_grid2}
\end{figure*}
We initialized all rotating models  to have solid body rotation at a fraction of the critical velocity $\omega_{\rm{i}} = \Omega_{\rm{i}} / \Omega_{\rm{crit}}$ just before the zero age main sequence (ZAMS). 
The effects of rotation on the stellar structure equations are included in \verb|MESA| using the \emph{shellular approximation} \citep[][]{Kippenhahn&Thomas1970, Endal&Sofia1976, Meynet&Maeder1997, Heger+2000, Paxton+2013, Paxton+2019}, for which \verb|MESA| solves the stellar structure equations in one dimension by projecting them along the isobars, with shape determined by the Roche potential \citep[][]{Maeder2009, Paxton+2019}. \verb|MESA| employs analytical fits to the Roche potential of a point mass, enabling the calculation of the structure of rotating stars up to $\omega \approx 0.9$ \citep[][]{Paxton+2019}.

Different definitions of the critical rotational velocity \citep[][]{Rivinius+2013} are used in various stellar evolution codes. \verb|MESA| adopts the following definition \citep[][]{Paxton+2013}:
\begin{equation}
    \Omega_{\rm{crit, MESA}}^2 = \left(1 - \frac{L}{L_{\rm{Edd}}}\right) \frac{GM}{R^3},
    \label{omega_crit_MESA}
\end{equation}
where $L$ is the total luminosity, $M$ is the total mass, $G$ is the universal gravitational constant, $R$ is the total radius, neglecting deviations from spherical symmetry, and $L_{\rm{Edd}} = 4 \pi c GM / \kappa$ is calculated as a mass-weighted average in a user-specified optical depth range (with the default value $\tau \in [1\text{--}100]$). Here, $c$ is the speed of light, and $\kappa$ is the Rosseland mean opacity. For $L / L_{\rm{Edd}} \sim 0$, as in our case, the critical angular and linear velocities are then defined as:
\begin{equation}
    \Omega_{\rm{crit,MESA}} = \sqrt{\frac{GM}{R_e^3}} \quad \text{and} \quad v_{\rm{crit,MESA}} = \sqrt{\frac{GM}{R_e}},
    \label{v_crit_MESA}
\end{equation}
where $v_{\rm{crit,MESA}}$ is the orbital velocity at the equator, $\Omega_{\rm{crit,MESA}}$ is the corresponding angular velocity and $R_{e}$ is the equatorial radius of the rotating star.
A more physically motivated definition in the context of this study is the critical velocity in the Roche model  \citep[][]{Maeder2009}:
\begin{equation}
    \Omega_{\text{crit}} = \sqrt{\frac{GM}{R^3_{\text{e,crit}}}} = \sqrt{\frac{8}{27} \frac{GM}{R^3_{\text{p,crit}}}}.
    \label{omega_crit_roche}
\end{equation}
In this framework, the critical linear velocity at the equator is defined as:
\begin{equation}
    v_{\text{crit}} = \sqrt{\frac{GM}{R_{\text{e,crit}}}} = \sqrt{\frac{2}{3} \frac{GM}{R_{\text{p,crit}}}},
    \label{v_crit_roche}
\end{equation}
noting that in this case $\Omega / \Omega_{\rm{crit}} \neq v / v_{\rm{crit}}$.
Here, $R_{\rm{e,crit}}$ and $R_{\rm{p,crit}}$ are the equatorial and polar radii of a critically rotating star, respectively\footnote{The variation of the polar radius $R_p$ due to rotation is small, so $R_{\rm{p,crit}} / R_p \approx 1$ is a reasonable approximation \citep[][]{Ekstroem+2008}. Therefore, Equation~\ref{v_crit_roche} and \ref{omega_crit_roche} can be calculated without requiring a numerical model of a critically rotating star.}. For more details, see \citet{Maeder2009}.
This definition is widely used in studies of stellar rotation statistics \citep[][]{Royer+2007} and is adopted by several stellar evolution codes, such as \verb|GENEC| \citep[][]{Eggenberger+2008} and \verb|PARSEC| \citep[][]{Nguyen+2022}.

For stellar models with identical input physics, mass, chemical composition, and initial $\omega$, but with different definitions of the critical velocity, the model using Equation~\ref{omega_crit_MESA} exhibits equatorial velocities, $v_{\rm{eq}}$, approximately 1.6 times larger than using Equation~\ref{omega_crit_roche}.
Following \citet{Wang+2023}, we convert $W := \Omega / \Omega_{\rm{crit,MESA}}$ to $\omega := \Omega / \Omega_{\rm{crit}}$, and vice versa, using Equation~11 in \citet{Rivinius+2013} and its inverse:
\begin{equation}
    \omega = \left(\frac{3 W^{2/3}}{2 + W^2}\right)^{3/2}.
\end{equation}
Note that $W = \omega$ only for non-rotating and critically rotating stars.

To account for the latitudinal variation of surface temperature and flux, we adopt the gravity darkening model of \citet{Espinosa-Lara&Rieutord2011}, which applies at any rotation rate and agrees well with interferometric observations of fast rotators \citep[][]{Espinosa-Lara&Rieutord2011, vanBelle2012}.
The luminosity measured by an observer at inclination $i$ with respect to the rotation axis is:
\begin{equation}
L_{\mathrm{MES}}(i) = 4\pi \int_{\mathbf{d\Sigma \cdot d} > 0} I(\theta)\, \mathbf{d\Sigma \cdot d}, 
\label{eq:L_MES}
\end{equation}
where $I(\theta)$ is the specific intensity at colatitude $\theta$, integrated over the visible hemisphere. Assuming blackbody radiation, this becomes:
\begin{equation}
L_{\mathrm{MES}}(i) = 4\sigma \int_{\mathbf{d\Sigma \cdot d} > 0} T^4_{\mathrm{eff}}(\theta)\, \mathbf{d\Sigma \cdot d}.
\label{eq:L_MES_BB}
\end{equation}
The corresponding effective temperature inferred by the observer is
\begin{equation}
T_{\mathrm{eff,MES}} = \left( \frac{L_{\mathrm{MES}}}{\sigma \Sigma_p} \right)^{1/4},
\label{eq:Teff_MES}
\end{equation}
where $\Sigma_p$ is the projected surface area perpendicular to the line of sight, and $\sigma$ is the Stefan–Boltzmann constant.
This expression differs from the standard definition of the effective temperature,
\begin{equation}
T_{\mathrm{eff}} = \left( \frac{L}{\sigma \Sigma} \right)^{1/4},
\label{eq:Teff}
\end{equation}
where $\Sigma$ is the total surface area of the star.

Another important aspect of rotation, particularly for stars with masses below approximately $1.5~\text{M}_{\odot}$, is magnetic braking \citep[][]{Kraft1967}. Low-mass dwarfs rapidly lose their initial angular momentum due to magnetized stellar winds, a process that is both theoretically expected and observed in the distribution of rotational velocities in star clusters and field stars \citep[][]{Beyer&White2024}.
However, rotation and magnetic activity are related to stellar age, mass and metallicity through a complex relationship that remains not fully understood. With precise stellar ages from asteroseismology, the relationship between age, rotation, mass and magnetic activity requires a ``weakened magnetic braking'' to explain the observations \citep[e.g.][]{vanSaders+2016,Metcalfe+2020}.
In the absence of a detailed magnetic braking model, we adopt a simplified approach to approximate its effects. 
Below a certain mass threshold, that depends on metallicity, we force models to be non-rotating. 
We limit the maximum $\omega_{\rm{i}}$ (see Section~\ref{subsubsec:rotation} for definition) that the stars can have for each stellar mass between $M_{\rm{norot}}$ and $M_{\rm{rot}}$, and this maximum increases with mass from $M_{\rm{norot}}$ to $M_{\rm{rot}}$ where the full range of rotation rates is computed. 
For example, for models with $Z=0.01$ we set $M_{\rm{norot}} = 1.3~\text{M}_{\odot}$ and $M_{\rm{rot}} = 1.6~\text{M}_{\odot}$; and for models with $Z=0.004$, we use $M_{\rm{norot}} = 1.2~\text{M}_{\odot}$ and $M_{\rm{rot}} = 1.5~\text{M}_{\odot}$. The decrease of $M_{\rm{norot}}$ and $M_{\rm{rot}}$ with metallicity is included to roughly account for the findings of \citet{Amard&Matt2020}, which show that metal-poor stars experience less efficient spin-down at ages $\gtrsim 1$ Gyr. A similar approach is employed in \citet{Gossage+2019} to mimic magnetic braking in low-mass stars in synthetic star clusters.

\subsubsection{Mass loss}\label{subsubsec:mass_loss}
For slow rotators in the range of masses we covered, a negligible amount of mass loss occurs throughout the MS. For a fast rotator however, it is possible to reach the critical rotation velocity during the MS, even if it started on the ZAMS with $\omega_{\rm{i}}<1$. Once critical rotation is attained, the effective gravity at the equator of the star vanishes. We therefore expect a strong enhancement of the mass-loss in the equatorial region, which removes overcritical layers and maintains the surface at, or slightly below,  critical rotation velocity. A description of this specific process is not yet included in stellar evolution codes, and numerically it is not yet possible to maintain the model at the critical velocity. To overcome this, we used a radiative wind prescription based on \citet{deJager+1988} with a rotationally-enhanced mass loss \citep[as in ][]{Paxton+2013}. 
Rotationally enhanced mass loss is adjusted iteratively before the next time step until the rotation rate at the surface is slightly subcritical  ($\omega_{\rm{max}}=0.99$). In effect, this is an implicit solution for the mass loss (of the order of 0.1\% of the total mass for the models with $\omega_{\rm{i}}=0.9$ and a much smaller fraction for slower rotators) when rotation is close to becoming super-critical \citep[][]{Paxton+2013}.

For the age range we considered (0.8-1.5~Gyr), asteroseismic measurements of mass for stars in open clusters suggests very little or null red giant branch (RGB) mass loss \citep[e.g.][]{Handberg+2017}. Therefore we did not include mass loss on the RGB in our models, but did include mass loss with $\eta_{B}=0.2$ for the asymptotic giant branch \citep{Bloecker1995}.

\subsubsection{Rotation induced transport of angular momentum and chemical elements}\label{subsubsec:rot_mix}
In line with most available stellar model grids including rotation, we consider only hydrodynamical transport processes for angular momentum and chemical elements \citep[][]{Georgy+2013, Choi+2016, Nguyen+2022}.
In \verb|grid 1|, we adopted the default \verb'MESA' fully diffusive formulation of rotation-induced mixing, as described in \citet{Heger+2000}. This formulation considers five rotationally induced hydrodynamical instabilities: dynamical shear instability, secular shear instability, Solberg-Høiland instability, Eddington-Sweet circulation, and Goldreich-Schubert-Fricke instability. There are two free parameters in this implementation: $f_{c}$, a number between 0 and 1 that represents the ratio of the diffusion coefficient to the turbulent viscosity; and $f_{\mu}$, that encodes the sensitivity of rotational mixing to the mean molecular weight gradient, $\nabla_{\mu}$ \citep[][]{Heger+2000, Paxton+2013}.
We chose  $f_{c} = 1/30$ and $f_{\mu} = 0.05$ following \citet{Pinsonneault+1989, Heger+2000}.

In \verb|grid 2|, we adopted a custom rotational mixing prescription during the MS, calibrated to emulate the transport of chemical elements and the evolution of surface equatorial velocity in the \verb|GENEC| models of \citet{Georgy+2013}.
The calibration of the diffusion coefficients for this prescription is described in Appendix~\ref{app:rot_mix_calib}. \verb|GENEC| adopts an advective-diffusive approach that includes the effects of meridional circulation and shear instabilities. The diffusion coefficients are derived within a self-consistent framework, under certain assumptions, and require the calibration of a single free parameter \citep[][]{Eggenberger+2008, Maeder2009, Georgy+2013}. In the post-main sequence, we switch to the default \verb|MESA| mixing\footnote{Note that {\ttfamily GENEC} models also switch to a fully diffusive transport of angular momentum and chemical elements in the post-MS phases \citep[][]{Nandal+2024}.}.  
This approach for \verb|grid 2| allowed us to approximate the rotation-induced mixing of chemical elements from \verb|GENEC| models within \verb|MESA| (see Appendix \ref{app:rot_mix_calib}), removing discrepancies due to differing input physics and assumptions between codes, and isolating differences in internal mixing efficiency between the two model grids.
Importantly, this hybrid method enabled the computation of the helium-burning phase in stars with masses below 2.5~M$_{\odot}$, outside of the \verb|GENEC| computational regime.
Furthermore, we extended \verb|grid 2| to masses below the 1.7~\Msun{} limit of \citet{Georgy+2014}, by linearly extrapolating the calibrated diffusion coefficients (as described in Appendix~\ref{app:rot_mix_calib}) to match the mass limits with \verb'grid 1', but we must refrain from drawing conclusions in this mass range when using \verb'grid 2'. 

At the start of the MS, rotational effects are nearly identical in both grids (see left panel of Figure~\ref{fig:grid1_vs_grid2}), with rotating models seen equator-on ($i=90^\circ$) showing lower measured luminosities and effective temperatures due to gravity darkening. For pole-on views ($i = 0^\circ$), the measured luminosity increases moderately, and the effective temperature shifts slightly towards higher values. However, as the star evolves on the main sequence, differences in the efficiency of rotational mixing become increasingly significant. In \verb|grid 2|, rotational mixing more efficiently supplies fresh hydrogen fuel to the core and transports helium and other H-burning products into the radiative zone.  
By increasing the size of the convective core and altering the chemical composition profiles in the radiative zone, rotational mixing induces an increase in luminosity and a widening of the main sequence \citep[discussed in detail in][]{Eggenberger+2011}. The models of \verb|grid 1| (top right panel of Figure \ref{fig:grid1_vs_grid2}) exhibit a significantly smaller change in the nitrogen-to-hydrogen ratio compared to the models of \verb|grid 2| (bottom).

\verb|grid 1| and \verb|grid 2| represent two commonly used approaches for implementing rotation-induced mixing in stellar models. The first is adopted in grids like \verb|MIST| \citep{Choi+2016, Gossage+2019}, the second in \verb|GENEC| \citep{Georgy+2013}. We attempt, for the first time, to constrain the impact of these different prescriptions using observations of intermediate-age star clusters with rotationally extended MSTOs.
% End of Grids of Stellar Models with Rotation

\section{Synthetic clusters} \label{sec:iso_clust_construction}
We computed synthetic clusters and isochrones from \verb|grid 1| and \verb|grid 2| by interpolating the evolutionary tracks using the \verb|SYCLIST| code \citep[see][for details]{Georgy+2014}.  
We describe our method, which follows the \emph{equivalent evolutionary points} (EEPs) approach described in \citet{Dotter2016} closely, and compare the resulting isochrones with those from other databases in Appendix \ref{track_sampling}.

We created a basis of single age synthetic stellar populations each with a uniform distribution of $\omega_i$ in a $\Delta\omega_i=0.1$ bin, centred at $\omega_i = 0.05, 0.15, 0.25, 0.35, 0.45, 0.55, 0.65, 0.75, 0.85, 0.95$. Each \textit{base stellar population} (BSP) contains $1.5 \times 10^5$ stars with mass between 0.6 \Msun{} - 4.0 \Msun{} adopting a \citet{Salpeter1954} initial mass function. BSPs were computed with ages ranging from $\log(\rm{age/yr})=8.900$ to $\log(\rm{age/yr})=9.180$, spaced by 0.005~dex in age.
We assumed a uniformly distributed orientation of the rotation axis which corresponds to a $\sin i$ probability distribution for the viewing angle  $i$.
Gravity and limb darkening were treated following \citet{Espinosa-Lara&Rieutord2011} and \citet{Howarth2011}, respectively. For implementation details, see \citet{Georgy+2014}.
BSPs were computed adopting a binary fraction of 0.2, roughly consistent with the binary fractions found for NGC~419 \citep[][]{Rubele+2010} and NGC~1817 \citep[][]{deJuanOvelar+2019} star clusters. The members of binary systems were drawn from the same grids of single stellar models and treated as unresolved binaries in the synthetic CMD as in \citet{Hurley&Tout1998}.

We found the best-fit synthetic clusters for both NGC~419 (Section \ref{subsec:NGC419}) and NGC~1817 (Section \ref{subsec:NGC1817}) by searching the optimal age and combination of BSPs that best reproduce the distribution of stars on the eMSTO and eRC of these clusters. 
As demonstrated in Section \ref{sec:rot_mix_eMSTOs_eRCs}, these two features are the most sensitive to the effects of rotation, rotational mixing and age.
Moreover the field contamination for these CMD features is negligible in both the studied clusters \citep[][]{Girardi+2009, Sandquist+2020}.
\begin{figure}
    \centering
	\includegraphics[width=\columnwidth]{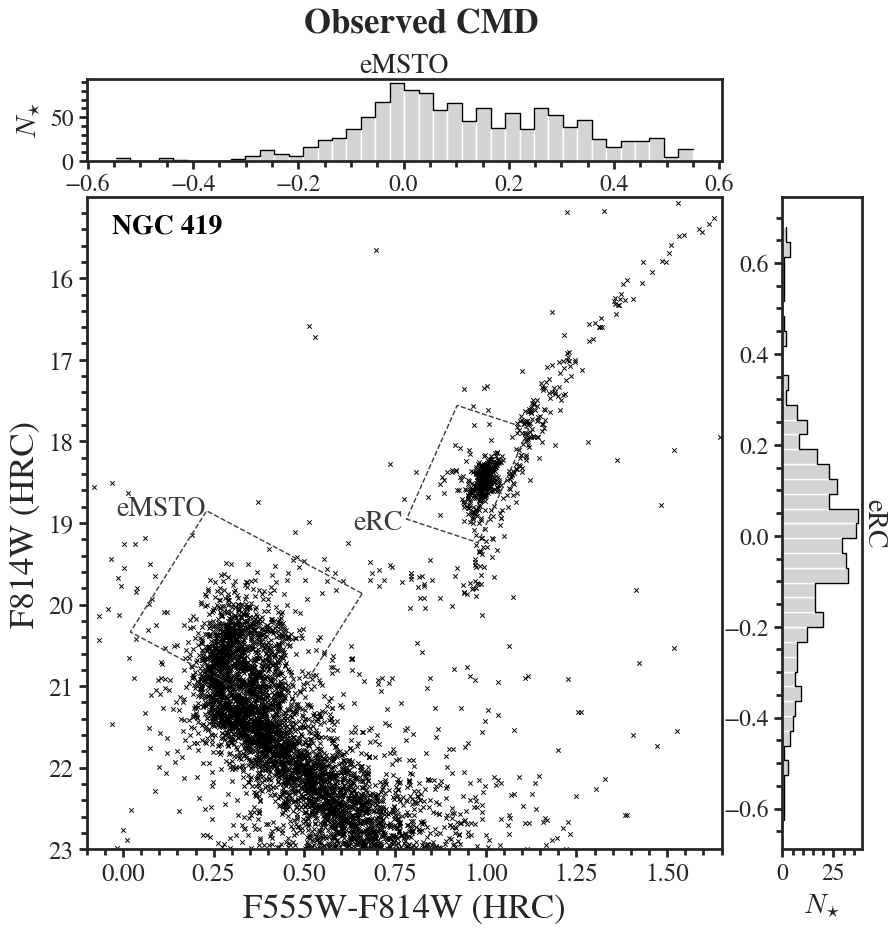}
    \caption{Observed CMD of NGC~419 from the HST programme GO-10396 (PI: J.~S.~Gallagher), specifically the ACS/HRC data were reduced by \citet{Girardi+2009}. The dashed parallelograms are the eMSTO and eRC regions described in Section \ref{sec:iso_clust_construction}. The top panel represents the histogram of star counts projected along the long axis of the eMSTO parallelogram, while the right panel represents the histogram of star counts projected along the long axis of the eRC parallelogram.}
    \label{fig:NGC419_data}
\end{figure}
We defined a parallelogram encompassing the eMSTO (see Figs. \ref{fig:NGC419_data} and \ref{fig:NGC1817_data}), with its long axis aligned along the direction in which gravity darkening extends the MSTO. This direction is determined by connecting the reddest and bluest points of two isochrones with $\omega_{\rm i}=0.9$, viewed equator-on and pole-on, respectively. For this preliminary step, the isochrone ages and metallicities were adopted from previous literature estimates for each cluster \citep[][]{Sandquist+2020, Ettorre+2025}. The size of the eMSTO parallelogram was chosen to enclose both the observed eMSTO and the MSTOs of all BSPs across the full age grid. We excluded the Kraft break region from this selection, since our models do not include a reliable treatment of magnetic braking and its inclusion could bias the fitting procedure. A second parallelogram was defined to encompass the eRC (see Figs. \ref{fig:NGC419_data} and \ref{fig:NGC1817_data}), with its long axis aligned with the observed extension of the eRC. This direction was determined by connecting the brightest and faintest points of the eRC. The size of the eRC parallelogram was set to best isolate the observed eRC from the RGB and early-AGB, while still enclosing the RC of all BSPs across the age grid.
\begin{figure}
    \centering
	\includegraphics[width=\columnwidth]{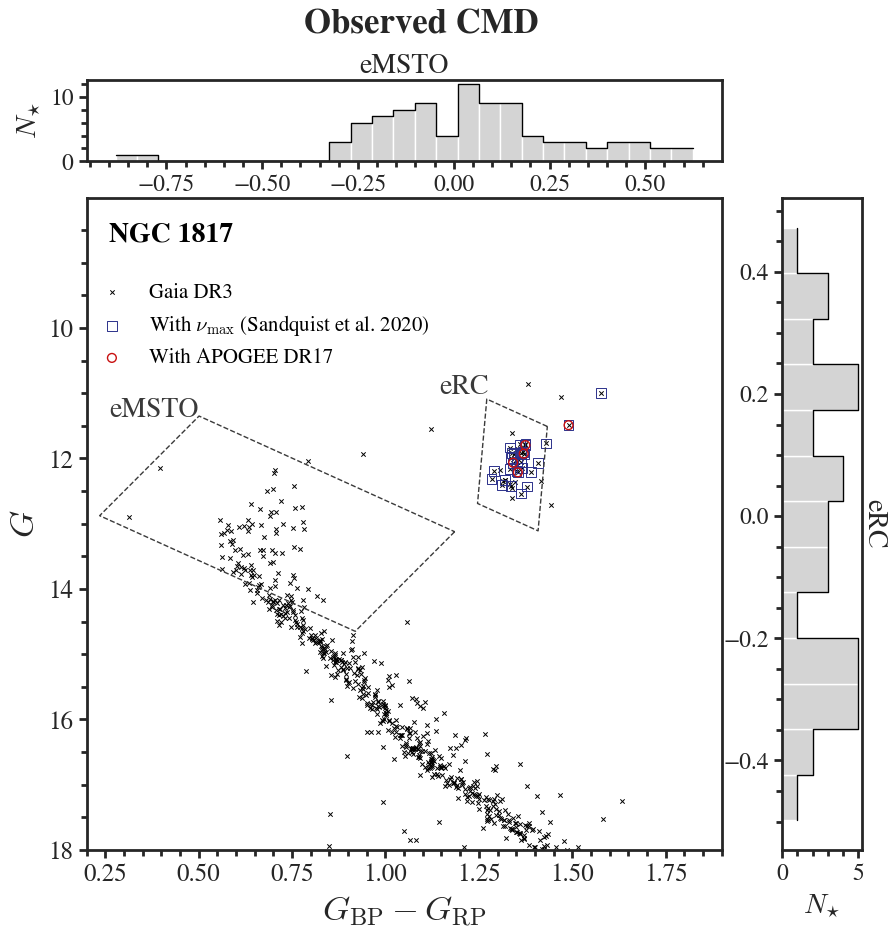}
    \caption{Observed CMD of NGC~1817. The stars enclosed in the blue boxes have $\nu_{\rm{max}}$ measured by \citet{Sandquist+2020} from K2 data. The stars encircled in red have also APOGEE DR17 spectroscopic parameters \citep[][submitted]{Willett+2025submitted}. The dashed parallelograms and the top and right panels are defined in the same way as Fig. \ref{fig:NGC419_data}.
    }
    \label{fig:NGC1817_data}
\end{figure}
We projected the coordinates of the stars within the eMSTO and eRC parallelograms along their respective long axes and constructed histograms of star counts for both the observations and the synthetic clusters (top and right panels of Figures~\ref{fig:NGC419_data} and \ref{fig:NGC1817_data}). The best-fit synthetic cluster was identified by determining the optimal combination of BSPs and age that minimizes the sum of the negative logarithmic Poisson likelihood ratios of the eMSTO and eRC histograms: 
\begin{equation}
    -\ln \mathcal{L}_{\rm P} = \sum_{i=\rm{eMSTO}}^{\rm{eRC}} \sum_{k=1}^{N_{\rm{bins}}} m_{k}^{i} - n_{k}^{i} + n_{k}^{i} \ln \frac{n_{k}^{i}}{m_{k}^{i}}
    \label{PLR},
\end{equation}
where $k$ indexes the histogram bins in each region $i$. This statistic is particularly suited to our analysis since it properly accounts for the discrete and Poissonian nature of star counts \citep[][]{Dolphin2002, Cordoni+2022, Ettorre+2025}.

The different BSPs are multiplied by a scaling factor, $C_j$, which specifies the contribution of each BSP to the final synthetic cluster before being combined. The coefficients $C_j$ can take values between 0 and 1. To minimize $-\ln \mathcal{L}_{\rm P}$ and determine the optimal set of $C_j$ and the cluster age, we employed the genetic algorithm implemented in the \texttt{PyGAD} Python library \citep[][]{PyGAD}, which reduces the risk of converging to a local minimum and facilitates the search for the global solution.
% End of Synthetic Clusters

\section{The impact of rotational mixing on extended main sequence turn-offs and extended red clumps}\label{sec:rot_mix_eMSTOs_eRCs}

\begin{figure*}
    	\includegraphics[width=\textwidth]{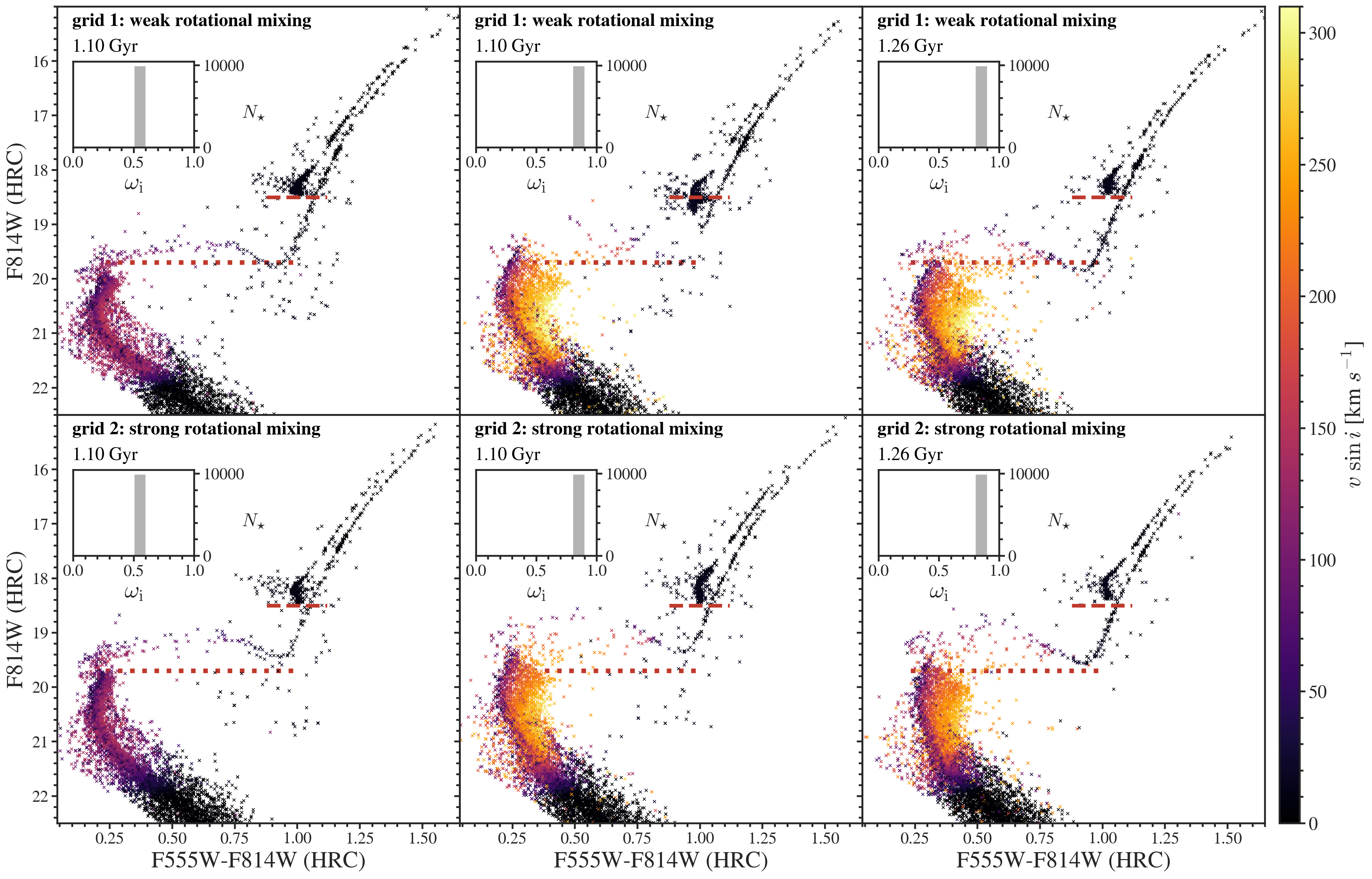}
    \caption{CMDs of selected BSPs at different rotation rates and different ages, for both \texttt{grid 1} (top panels) and \texttt{grid 2} (bottom panels), transformed into the observational plane of NGC 419 as described in Section \ref{subsec:NGC419}. Going from left to right, for both the top and bottom panels, the following BSPs are represented: $\sim1.10$~Gyr ($\log(\rm{age/yr})=9.040$) centered at $\omega_{i}=0.55$, $\sim1.10$~Gyr centered at $\omega_{i}=0.85$, $\sim1.26$~Gyr ($\log(\rm{age/yr})=9.100$) centered at $\omega_{i}=0.85$. The color map represents the projected rotational velocity $v \sin i$ of the stars. For representation purposes we selected only a random subsample of $10^4$ stars from the original BSPs. The inset histogram shows the distribution of initial rotation rates $\omega_{i}$ of the stars in the BSP. For reference, the dashed and dotted red lines mark the eRC magnitude of peak density and the average SGB magnitude of NGC~419, respectively.}

    \label{fig:base_synth_pops}
\end{figure*}
A selection of BSPs centered at different rotation rates and ages, for both \texttt{grid 1} (weak rotational mixing) and \texttt{grid 2} (strong rotational mixing), is shown in Fig.~\ref{fig:base_synth_pops}. These BSPs have been transformed into the observational plane of NGC~419 as described in Section~\ref{subsec:NGC419}. For visualization purposes, we display only a random subsample of $10^4$ stars from the original BSPs.

The top panels of Fig.~\ref{fig:base_synth_pops} show three BSPs from \texttt{grid 1}. For reference, the dashed and dotted red lines mark the eRC magnitude of peak density and the average SGB magnitude of NGC~419, respectively. From top-left to top-right: the first two correspond to the same age of $\sim1.10$~Gyr ($\log(\rm{age/yr})=9.040$) but are centered at different rotation rates, $\omega_{i}=0.55$ and $\omega_{i}=0.85$, respectively; the third BSP corresponds to an age of $\sim1.26$~Gyr ($\log(\rm{age/yr})=9.100$) and is centered at $\omega_{i}=0.85$. Four clear effects emerge as the average $\omega_{i}$ and age increases: 
\begin{enumerate}
    \item Gravity darkening broadens the MSTO region: stars observed at $i \sim 90^\circ$ (large $v \sin i$) appear redder and fainter, while those at $i \sim 0^\circ$ (small $v \sin i$) appear slightly bluer and brighter.
    \item Gravity darkening also widens the early portion of the SGB, which then narrows toward the base of the RGB as stellar rotation decreases with envelope expansion.
    \item At an age of $\sim1.10$~Gyr, rotational mixing extends the RC to fainter magnitudes and slightly bluer colours. This effect arises from rotational mixing triggered by fast rotation during the MS phase, rather than from the stars’ current rotational velocities. In the case of the eRC of NGC~419, \citet{Dresbach+2023} argued that the slow rotation of RC stars rules out significant rotational effects, but this interpretation overlooks the role of rotational mixing during the MS. With an average $\omega_{i}=0.55$, only the compact primary RC, populated by stars that ignited helium in fully degenerate cores and experienced the He flash, is present. In contrast, with an average $\omega_{i}=0.85$, both the primary and secondary clumps appear. Even the weak rotational mixing of \texttt{grid 1} models at $\sim1.10$~Gyr is sufficient to slightly increase the He-core mass, allowing the initially fastest rotators to avoid the He flash by reaching helium ignition before the core becomes fully degenerate, thus populating the secondary clump. 
    \item At an age of $\sim1.26$ Gyr and an average $\omega_{i}=0.85$, (i) and (ii) remain valid as in the $\sim1.10$ Gyr case, whereas (iii) no longer applies. At this older age, rotational mixing is no longer sufficient to increase the He-core mass beyond the threshold required to avoid helium ignition in fully degenerate cores, so all stars undergo the He flash and end up in the compact primary RC. This naturally explains the case of clusters that display an eMSTO but a compact RC \citep[e.g.][]{Li+2014a, Li+2014b, Li+2016b, Li+2024}.
\end{enumerate}
It is interesting to note that only the case with an eRC, at an age of $\sim1.10$~Gyr and centered at $\omega_{i}=0.85$, matches the magnitude of peak density of the eRC in NGC~419 (red dashed line in Fig.~\ref{fig:base_synth_pops}). This is consistent with the interpretation of the NGC~419 eRC as a combination of primary and secondary RC populations \citep[][]{Girardi+2009}.

The bottom panels of Fig.~\ref{fig:base_synth_pops} show three BSPs corresponding to those in the top panels but computed from \texttt{grid 2} models. Considerations (i), (ii), (iii) and (iv) made for \texttt{grid 1} still apply, but with three key differences:
\begin{enumerate}
    \item With an average $\omega_{i}=0.85$, the post-MS phases appear $\sim0.5$~mag brighter and slightly bluer than in the case of weak rotational mixing from \texttt{grid 1}. This is a consequence of the strong rotational mixing in \texttt{grid 2} models, which transports helium and other H-burning products into the radiative zone and significantly increases the size of the convective core. The resulting higher mean molecular weight in the envelope and larger He-core mass lead to brighter post-MS phases \citep[][]{Maeder2009, Eggenberger+2011, Salaris&Cassisi2017}.
    \item For the same average $\omega_{i}$, the eMSTO is less extended compared to the case of \texttt{grid 1}. This results from the strong rotational mixing in \texttt{grid 2} models, which counteracts the effect of gravity darkening. While gravity darkening shifts stars observed nearly equator-on toward redder colors and fainter magnitudes, rotational mixing tends to move stars toward bluer colors and brighter magnitudes, regardless of their inclination to the line of sight \citep[][]{Girardi+2011}.
    \item For the same average $\omega_{i}$, the maximum $v \sin i$ of eMSTO stars is lower in the \texttt{grid 2} case.
\end{enumerate}
It is interesting to note that none of the BSPs computed with \texttt{grid 2} models match the magnitude of peak density of the eRC (red dashed line in Fig.~\ref{fig:base_synth_pops}) or the average magnitude of the SGB (red dotted line in Fig.~\ref{fig:base_synth_pops}) of NGC~419.

\begin{figure}
    \centering
	\includegraphics[width=\columnwidth]{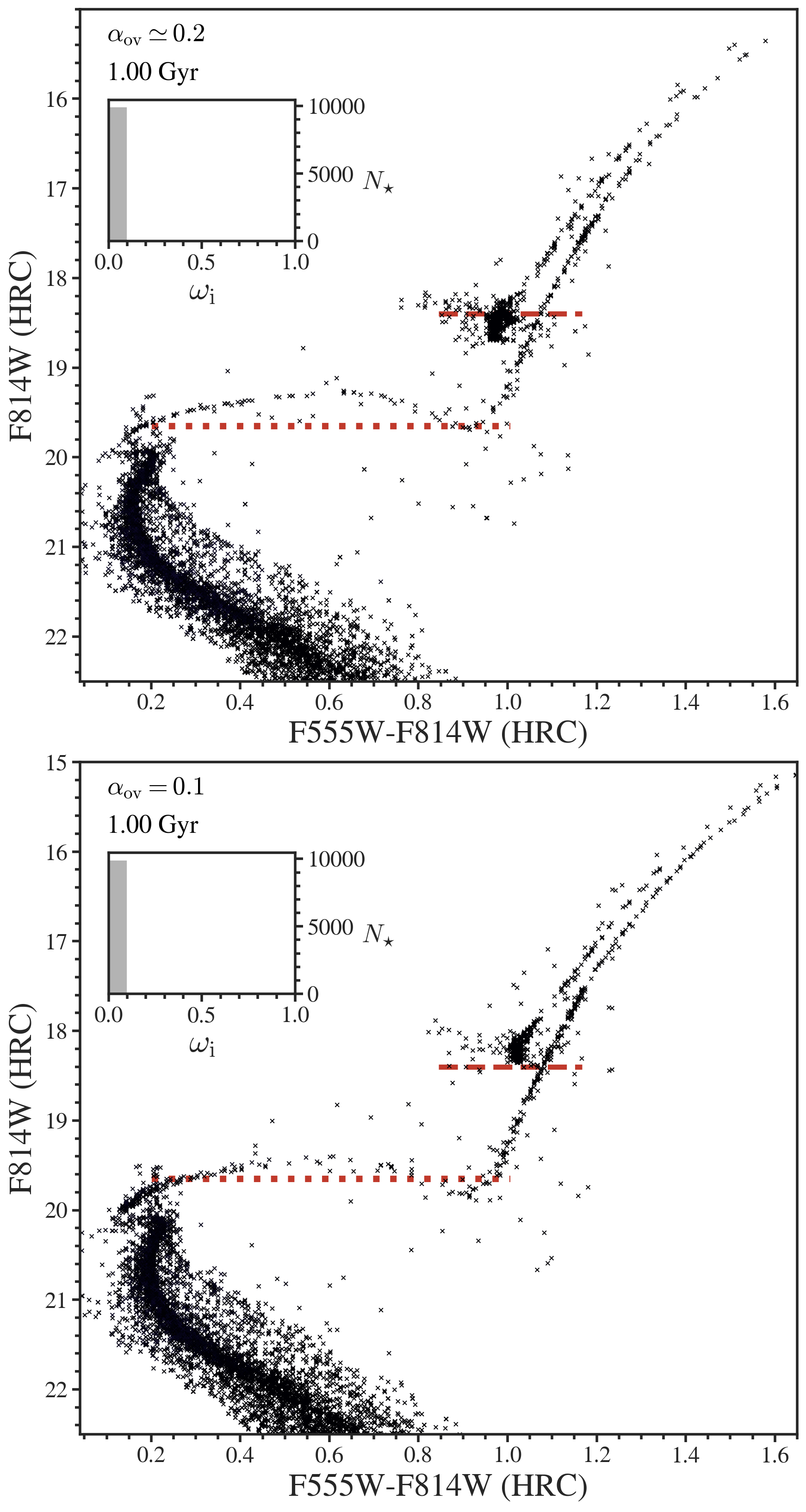}
    \caption{CMDs of two BSPs with an age of 1 Gyr centered at $\omega_i=0.05$ (inset histogram) with different amount of convective core overshooting: $\alpha_{\rm {ov}}\simeq0.2$ (top panel) and $\alpha_{\rm {ov}}=0.1$ (bottom panel) transformed into the observational plane of NGC 419 as described in Section \ref{subsec:NGC419}. For representation purposes we selected only a random subsample of $10^4$ stars from the original BSPs. For reference, the dashed and dotted red lines mark the eRC magnitude of peak density and the average SGB magnitude of NGC~419, respectively.}
    \label{fig:ovsh_eRC}
\end{figure}

It must be acknowledged that the amount of convective core overshooting during the MS can affect, in a way similar (though not identical) to rotational mixing, the magnitudes of the post-MS phases \citep[][]{Eggenberger+2010} and influence whether a cluster exhibits an eRC or a compact RC at a given age. Fig.
~\ref{fig:ovsh_eRC} shows two BSPs with a central $\omega_{i}=0.05$ (essentially non-rotating) at an age of 1.0~Gyr, but with different amounts of convective core overshooting. At this age, the BSP with larger overshooting (top panel of Fig.~\ref{fig:ovsh_eRC}) exhibits an eRC with both primary and secondary clump and a slightly brighter SGB, whereas the BSP with smaller overshooting (bottom panel of Fig.~\ref{fig:ovsh_eRC}) displays a compact RC (primary RC only) and a slightly fainter SGB.

We selected two representative star clusters for modelling: the Small Magellanic Cloud (SMC) cluster NGC~419 and the Milky Way open cluster NGC~1817. Both clusters exhibit prominent eMSTOs and eRCs \citep[][]{Girardi+2009, Sandquist+2020} and are known to host fast-rotating stars on the MSTO. In Sections~\ref{subsec:NGC419} and~\ref{subsec:NGC1817}, we present the best-fit synthetic clusters for NGC~419 and NGC~1817, constructed using stellar models from \verb|grid 1| and \verb|grid 2|, as described in Section~\ref{sec:iso_clust_construction}.

\subsection{NGC~419}\label{subsec:NGC419}
NGC419 is an intermediate-age \citep[$\sim1.26$~Gyr;][]{Ettorre+2025} globular cluster in the SMC, exhibiting one of the most extended eMSTOs and an eRC observed to date. This cluster is believed to host a secondary clump of helium-burning stars that are just massive enough to avoid full electron degeneracy in their hydrogen-exhausted cores, alongside slightly less massive stars that ignited helium in a degenerate core \citep[][]{Girardi+2009}. Additionally, \citet{Kamann+2018} found a range of $v\sin i$ values between the blue and red MSTO stars in this cluster (see Section~\ref{subsubsect:NGC419_rotation}).

\begin{figure*}
    \centering
    \subfigure[]{\includegraphics[width=\columnwidth]{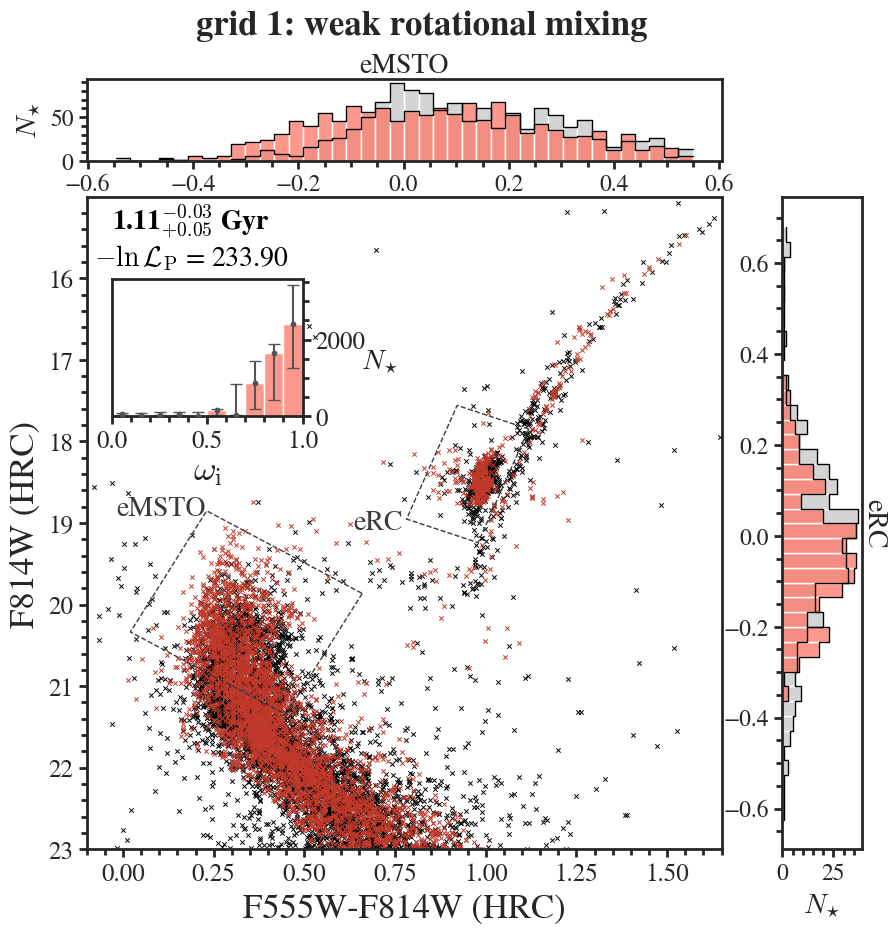}\label{fig:best_fit_NGC419_grid1}} 
    \subfigure[]{\includegraphics[width=\columnwidth]{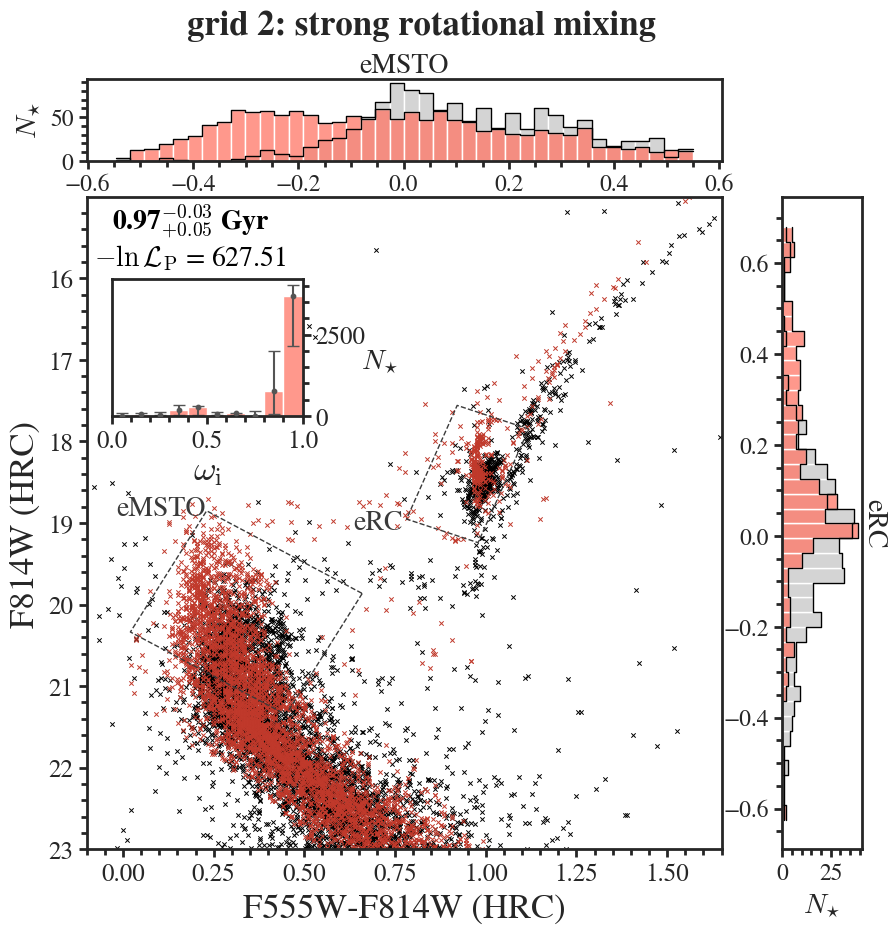}\label{fig:best_fit_NGC419_grid2}}
    \caption{
    Left: Best fit CMD (red) obtained combining BSPs from \texttt{grid 1} stellar models (weak rotational mixing) and the observed CMD of NGC~419 (black). Right: Best fit CMD (red) obtained combining BSPs from \texttt{grid 2} stellar models (strong rotational mixing) and observed CMD of NGC~419 (black). Each panel includes an inset with the distribution of initial stellar rotation rates $\omega_i$ for the corresponding synthetic cluster, along with the age of the best-fit synthetic cluster and the value of the negative logarithmic Poisson likelihood ratio $-\ln \mathcal{L}_{\rm P}$. The dashed parallelograms are the regions described in Section \ref{sec:iso_clust_construction} to study the distribution of stars across the eMSTO and eRC respectively. The histograms in the top and right panels show the star counts projected along the long axis of the eMSTO and eRC parallelograms, respectively. The best-fit synthetic cluster is shown in red, and the observations of NGC~419 are shown in black.
    }
    \label{fig:best_fits_NGC419}
\end{figure*}
We used the observations of NGC~419 from the HST programme GO-10396 (PI: J.~S.~Gallagher), specifically the ACS/HRC data were reduced by \citet{Girardi+2009}.
We adopted a metallicity of $Z = 0.004$, consistent with the value $\left[\mathrm{Fe}/\mathrm{H}\right] = -0.58 \pm 0.02$~dex derived for NGC~419 by \citet{Mucciarelli+2023}. Synthetic photometry was transformed into the observational plane using the \verb|YBC| bolometric corrections \citep{Chen+2019}, which include variable extinction coefficients for the \textit{Hubble Space Telescope} Advanced Camera for Surveys High Resolution Channel (ACS/HRC) F814W and F555W filters. We adopted the distance modulus $\left(m - M\right)_0 = 18.85 \pm 0.03$~mag from \citet{Goudfrooij+2014} and a colour excess $E(B-V) = 0.061 ^{+0.026}_{-0.033}$~mag from the reddening map by \citet{Skowron+2021}. We applied to the synthetic CMDs the photometric errors and incompleteness derived from artificial star tests performed on the original HRC images.

By simultaneously fitting the distributions of stars in the eMSTO and eRC (Section~\ref{sec:iso_clust_construction}), we obtained for \texttt{grid 1} a best-fitting age of $1.11^{-0.03}_{+0.05}$~Gyr, characterized by a dominant population of fast rotators ($\omega_{i}\gtrsim0.75$) and a smaller contribution from slower rotators ($\omega_{i}\simeq0.55$; see inset in Fig.~\ref{fig:best_fit_NGC419_grid1}).
A similarly high fraction of fast rotators in NGC~419 is also found in the best-fit model of \citet{Ettorre+2025}, although their derived age ($\sim1.26$~Gyr) is significantly older. This difference in best-fit age arises primarily from the different fitting approach, as they varied metallicity, distance, and foreground extinction while fitting the full CMD, whereas we fixed these parameters based on independent observations and focused on the eMSTO and eRC. In addition, differences in the adopted physics of the \texttt{PARSEC V2.0} models \citep[][]{Nguyen+2022}, including the implementation of rotational mixing as well as the mass and rotation rate resolution of the grids, which differ from both our \texttt{grid 1} and \texttt{grid 2}, likely contribute to the discrepancy in the derived ages.

For \texttt{grid 2} we found a best-fitting age of $0.97^{-0.03}_{+0.05}$~Gyr, a high fraction of fast rotators with $\omega_{i}\gtrsim0.85$ and a much smaller contribution of slower rotators with $\omega_{i}\simeq0.4$ (see inset in Fig. \ref{fig:best_fit_NGC419_grid2}).
The uncertainties were computed by repeating the optimization through 1000 Monte Carlo realizations, varying $E(B-V)$, $\left(m - M\right)_0$ within the uncertainties and bootstrapping with replacement the observed CMD of NGC 419. We neglected the contribution of the 0.02 dex uncertainty in $\left[\mathrm{Fe}/\mathrm{H}\right]$, as it produces negligible perturbations in the CMD compared to the uncertainties in $E(B-V)$ and $\left(m - M\right)_0$.

The best-fit synthetic cluster found using \texttt{grid 1} (weak rotational mixing) gives a $-\ln \mathcal{L}_{\rm P} = 233.90$, which is significantly smaller than that found for \texttt{grid 2} (strong rotational mixing case) that gives a $-\ln \mathcal{L}_{\rm P}=627.51$.

The inability to achieve a best-fit with \texttt{grid 2} comparable to that obtained with \texttt{grid 1} is due to the much stronger rotational mixing. As discussed in Section~\ref{sec:rot_mix_eMSTOs_eRCs}, in \texttt{grid 2} the strong rotational mixing systematically increases the brightness and slightly shifts all CMD features toward bluer colors, starting from the eMSTO. In addition, it slightly reduces the width of the eMSTO.

It is worth noting that the \texttt{grid 2} best-fit synthetic cluster is significantly younger than the \texttt{grid 1} best-fit. Although an older age would better reproduce the eMSTO, a younger age provides a better match to the faint extension of the eRC. This trend is also seen with \texttt{grid 1}, but in the case of \texttt{grid 2} the RC brightness is so strongly increased by rotational mixing that no satisfactory compromise can be achieved.

The best-fit synthetic cluster obtained with \texttt{grid 1}, although significantly better than that from \texttt{grid 2}, still shows discrepancies that suggest some ingredients in our rotating models are either poorly modelled or missing. The eMSTO in the best-fit cluster contains an excess of stars on the bright blue side, where pole-on fast rotators are expected. Even with a large fraction of near-critical rotators ($\omega_{i}\simeq0.95$), the faint red side of the eMSTO remains slightly less extended than observed. Similar limitations have been reported in other studies \citep{Goudfrooij+2017, Gossage+2019, Lipatov+2022}, where rotation alone could not reproduce the full eMSTO extension without invoking ad hoc assumptions on the orientation distribution of rapid rotators.

The SGB of the best-fit cluster is also on average slightly brighter than observed. While the eRC is broadly well reproduced in both magnitude and colour, it does not extend to magnitudes as faint as in the observation.

\subsubsection{Rotation across the eMSTO of NGC~419}\label{subsubsect:NGC419_rotation}
\begin{figure}
    \centering
	\includegraphics[width=\columnwidth]{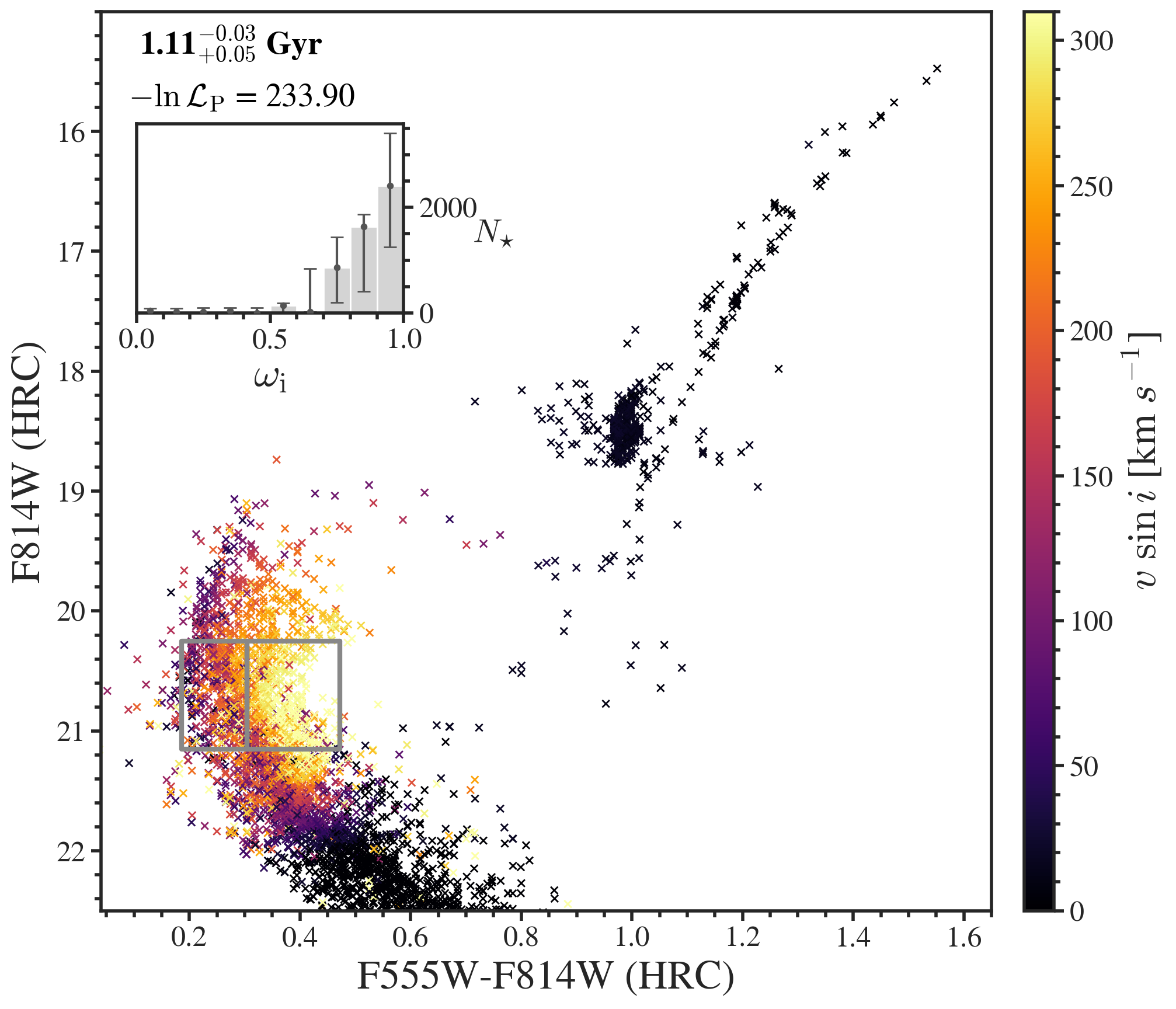}
    \caption{CMD of our best-fit synthetic cluster for NGC~419 from \texttt{grid 1} stellar models (left panel of Fig.~\ref{fig:best_fits_NGC419}), colored by $v\sin i$. The inset shows the histogram of initial rotation rates for the stars in the best-fit synthetic cluster. The gray boxes replicate the regions of the eMSTO where \citet{Kamann+2018} measure the average $v \ \sin i$ of NGC~419 stars. We obtained average $v\sin i$ values of $170 \pm 68$ and $258 \pm 62$~km~s$^{-1}$ for the blue and red MSTO regions, respectively.}
    \label{fig:NGC419_vsini_best}
\end{figure}
For NGC~419, \citet{Kamann+2018} derived average $v\sin i$ values of $87 \pm 16$ and $130 \pm 22$~km~s$^{-1}$ for stars in two regions spanning the blue and red MSTO, respectively. These values were obtained from combined spectra of stars selected within boxes that contain roughly equal numbers of stars on either side of the eMSTO. 
Figure~\ref{fig:NGC419_vsini_best} shows the distribution of $v\sin i$ across the CMD for our best-fit synthetic cluster of NGC~419 from \texttt{grid 1} (left panel of Fig.~\ref{fig:best_fits_NGC419}). The projected rotational velocities exhibit a clear trend, with lower values on the bluer side of the eMSTO and higher values on the redder side. This trend closely resembles the $v\sin i$ distribution observed across the eMSTO of NGC~1846 by \citet{Kamann+2020}.
Applying the same box definitions as \citet{Kamann+2018} to our synthetic CMD, we computed average $v\sin i$ values of $170 \pm 68$ and $258 \pm 62$~km~s$^{-1}$ for the blue and red MSTO regions, respectively.

In our synthetic cluster, the projected rotational velocities are systematically higher than the observed values of \citet{Kamann+2018}, with a larger difference between the blue and red MSTO regions.

This systematically larger $v \sin i$ in the synthetic clusters is also found by \citet{Bastian+2018, Kamann+2020, Kamann+2023}.
It is interesting to note that the possibility of stars in the synthetic cluster rotating too rapidly is consistent with the overabundance on the bright blue side of our best-fit synthetic cluster (Fig.~\ref{fig:best_fit_NGC419_grid1}), where pole-on fast rotators tend to be located.
The magnitude of this discrepancy, however, should be interpreted with caution, as measurement for almost critically rotating stars can be biased towards lower values because of the rotationally induced darkening of the equatorial regions of the star \citep[][]{Townsend+2004}. \citet{Kamann+2018} also note that systematic uncertainties in their MUSE measurements, due to the instrument’s relatively low spectral resolution, could reach up to 30~km~s$^{-1}$.  
Additionally any inaccuracy in the modelling of gravity darkening and bolometric corrections not specifically calibrated for rapidly rotating stars \citep[][]{Girardi+2019} could alter the best-fitting distribution of initial rotation rates required to reproduce the eMSTO morphology and, consequently, the predicted $v \sin i$ values across it.

An intriguing hypothesis is closely related to the phenomenon of Be \citep[][]{Milone+2018, Bastian+2016} and UV-dim stars \citep[][]{Martocchia+2023, Milone+2023b}. It is possible that a large fraction of the fast rotators in these clusters develop decretion discs when they reach high rotation rates, and that dust grains condense in the outer regions of these discs \citep[][]{D'Antona+2023, He+2025}. In clusters with ages around 1.0 Gyr or older, such discs may not be easily identifiable in emission, unlike for Be stars in younger clusters, since they are not hot enough to ionize the disc material. These stars could instead be identified in absorption \citep[][]{Kamann+2023}. The red side of the eMSTO could then be populated by moderately fast rotating stars whose dust rings intercept the line of sight, causing extinction of stellar light. The reddest UV-dim stars may represent the extreme tail of the distribution, corresponding to both maximum obscuration and orientations where the dust ring is aligned with the line of sight.
When only gravity darkening is considered, the eMSTO extension can be reproduced only by extremely fast rotators ($\omega_i>0.9$) observed close to equator-on. This would imply very high projected rotational velocities ($v \sin i \gtrsim 300$ km s$^{-1}$) that are not observed.

Instead, observations indicate that Be stars typically rotate at an average of $v/v_{\mathrm{crit}} \sim 0.7$ \citep[][]{Dufton+2022, Rivinius+2013, Kamann+2023}. If decretion discs can form at sub-critical rotation, through mechanisms that facilitate mass ejection such as the pulsationally driven orbital mass-ejection model \citep[][]{Kee+2016}, then stars rotating at lower $v/v_{\mathrm{crit}}$ could still be subject to extinction caused by the disc when viewed within the disc opening angle. In this case, the additional extinction may be sufficient to account for the observed width of the eMSTO without invoking extremely rapid rotators.

\begin{figure}
    \centering
	\includegraphics[width=\columnwidth]{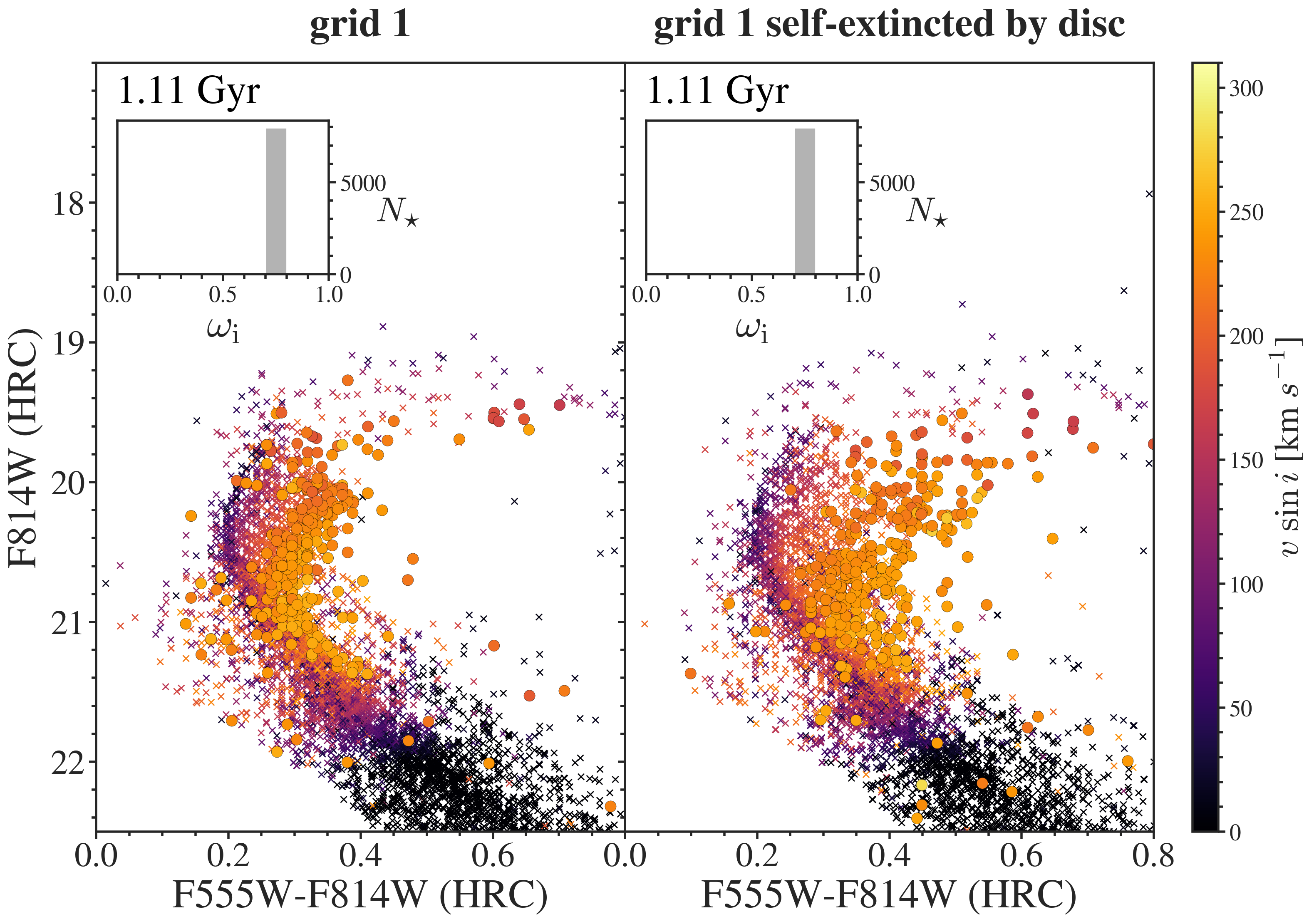}
    \caption{Right: CMD of a $\sim1.11$~Gyr ($\log(\rm{age/yr})=9.045$) old BSP from \texttt{grid 1} models centered at $\omega_i=0.75$ (inset histogram) and coloured by $v\sin i$. Left: CMD of the same BSP but including the contribution of the simple model of self-extinction by the decretion disc in stars observed within the disc opening angle and rotating faster than $v/v_{\rm{crit}} \gtrsim 0.7$ described in Section \ref{subsubsect:NGC419_rotation}.}
    \label{fig:NGC419_vsini_disc}
\end{figure}

We tested this idea with a simple experiment. Lacking a complete model of the disc structure and its wavelength- and angle-dependent optical depth, we explored the effect of increasing the reddening $E(B-V)$ for stars in our synthetic cluster rotating at $v/v_{\mathrm{crit}}\gtrsim0.7$ and observed within $\sim 15^{\rm{o}}$ of the equatorial plane. We modulate the reddening with a Gaussian dependence on the viewing angle $i$, and scale it with a logistic function of the normalized equatorial velocity:
\begin{equation} E(B-V) \rightarrow E(B-V)\left[1 + 2\cdot\mathcal{G}(i)\mathcal{S}(v_{\rm{eq}}/v_{\rm{crit}})\right], 
\label{eq:disc_extinction}\end{equation} 
where $\mathcal{G}(i)$ is a Gaussian with $\sigma_i = 15^{\circ}$ \citep[][]{Kamann+2023} centered at $\mu_i = 90^{\circ}$, and
\begin{equation} 
\mathcal{S}(x) = 1/\left[1 + \exp\left(-\tfrac{x-0.7}{0.08}\right)\right] 
\end{equation}
is a logistic factor that increases sharply for stars rotating faster than $\sim 70$\% of the critical velocity, and saturates such that the total reddening reaches up to three times the interstellar value. In absolute terms, this corresponds to a maximum additional extinction of $E(B-V) = 0.122$ mag from the disc. Figure \ref{fig:NGC419_vsini_disc} shows the CMD of a BSP at $\sim1.11$~Gyr ($\log(\mathrm{age/yr}) = 9.045$) with central $\omega_i = 0.75$, transformed into the observational plane as described in Section \ref{subsec:NGC419}. The maximum $v \sin i$ across the eMSTO is consistent with observations of intermediate-age clusters \citep[][]{Kamann+2018b, Kamann+2020}, but it is significantly lower than in our best-fit synthetic cluster (Fig. \ref{fig:NGC419_vsini_best}). Nevertheless, the MSTO remains relatively narrow. The right panel of Figure \ref{fig:NGC419_vsini_disc} shows the same CMD after applying the reddening scaling of Eq.~\ref{eq:disc_extinction} to account for self-extinction by the decretion disc in stars observed within the disc opening angle and rotating faster than $v/v_{\rm{crit}} \gtrsim 0.7$.
The effect broadens the eMSTO while maintaining realistic $v \sin i$ values and avoiding overpopulation of the blue-bright region by fast, pole-on rotators, unlike in our best-fit synthetic cluster (Figs. \ref{fig:best_fit_NGC419_grid1} and \ref{fig:NGC419_vsini_best}).
We emphasize that this experiment is highly schematic: the amount of extinction produced by the disc, its dependence on rotation, and the use of the interstellar extinction curve may not accurately reflect the true properties of disc-generated extinction. Decretion-disc formation could alter the stars’ angular momentum evolution and internal transport processes, including rotational mixing. This experiment therefore provides a zeroth-order test of the concept, demonstrating its potential and motivating further investigation. A detailed modeling of the disc structure and extinction curve is required and lies beyond the scope of this work. If physical models of decretion discs, including their optical depths, indicate a comparable impact in terms of self-extinction, this could significantly influence the inferred rotation distributions, cluster ages, and constraints on rotational mixing in clusters exhibiting an eMSTO.

\subsection{NGC~1817}\label{subsec:NGC1817}
NGC~1817 is a massive intermediate-age Galactic open cluster, approximately 1~Gyr old \citep[][]{Cordoni+2018, Sandquist+2020}, that exhibits a clear eMSTO and eRC. \citet{Sandquist+2020} suggest the presence of a subset of stars in the core helium-burning (CHeB) phase with core masses near the minimum required for non-degenerate helium ignition (i.e., secondary clump stars), along with more massive clump stars that have not yet evolved off the RC.

\citet{Sandquist+2020} also show that it is not possible to simultaneously reproduce the eMSTO and eRC using a single non-rotating isochrone. Analysing the eMSTO, \citet{Cordoni+2018} propose that stellar rotation, particularly the presence of stars rotating near critical velocity, may account for the observed morphology. This interpretation is consistent with the observation of \citet{Molenda-Zakowicz+2009}, who measured a range of $v\sin i$ values across the MSTO (see Section~\ref{subsubsect:NGC1817_rotation}).

Cluster members were identified by \citet{hunt+2024} using the \emph{Gaia}~DR3 catalogue \citep{GaiaDR32023}. Their photometry was then corrected for differential reddening following the procedure of \citet{Milone+2009} and adapted to open clusters by \citet{Cordoni+2018}.
First, we used the Gaia $G_\mathrm{BP}$ and $G_\mathrm{RP}$ reddening coefficients from \citet{Casagrande&VandenBerg2018} to define the reddening direction and rotate the $G_\mathrm{RP}$  vs. $G_\mathrm{BP} - G_\mathrm{RP}$ CMD, aligning the reddening vector with the new x-axis.
Next, we determined the fiducial line of the rotated CMD from a sample of reference stars, defined as bright main-sequence stars, excluding clear unresolved binary stars. Finally, we computed the distance, along the x-axis, of each reference star from the fiducial line. To calculate the differential reddening associated with each cluster star, we selected the 25 neighbouring reference stars and calculated the median distance along the reddening line. We excluded each reference star from the determination of its own differential reddening. The corresponding error is calculated as the root mean scatter of the distance values divided by $\sqrt{N - 1}$.

This cluster was observed during Campaign 13 of the \emph{K2} mission, and \citet{Sandquist+2020} measured average asteroseismic parameters ($\nu_{\rm{max}}$ and/or $\Delta\nu$) describing solar-like oscillation spectra of 29 giants, most of which were identified as RC stars. For five of these giants, spectroscopic chemical abundances and stellar parameters are available in the \citet[submitted]{Willett+2025submitted} catalogue, based on \emph{APOGEE}~DR17 spectra \citep[][]{Majewski+2017, Abdurro'uf+2022}.

For modelling NGC~1817 we adopted a metallicity of $Z~=~0.010$, consistent with the values $\left[\mathrm{Fe}/\mathrm{H}\right] = -0.08 \pm 0.02$ and $\left[\mathrm{Fe}/\mathrm{H}\right] = -0.11 \pm 0.03$ derived for NGC~1817 by \citet{Casamiquela+2017}. Synthetic photometry was transformed into the observational plane using the \verb|YBC| bolometric corrections \citep{Chen+2019}, which account for variable extinction coefficients in the \emph{Gaia}~DR3 $G$, $G_{\rm{BP}}$, and $G_{\rm{RP}}$ bands.
The distance modulus used to shift the synthetic cluster magnitudes was derived from the median Gaia~DR3 parallax of NGC~1817 stars. A zero-point correction was applied following \citet{Lindegren+2021}, using the relevant parameters from the Gaia~DR3 archive. This yielded a distance modulus of $\left(m-M\right)_0 = 11.101 \pm 0.008$. To estimate the color excess $E(B-V)$, we used five NGC~1817 giants with spectroscopic $T_{\rm{eff}}$ and $\log g$ from \emph{APOGEE}~DR17 spectra \citep[][submitted]{Willett+2025submitted}, and determined the value that best matched their \emph{Gaia}~DR3 $G_{\rm{BP}} - G_{\rm{RP}}$ colors using the \verb|YBC| bolometric corrections. This procedure yielded $E(B-V) = 0.228 \pm 0.008$.

\begin{figure*}
    \centering
    \subfigure[]{\includegraphics[width=\columnwidth]{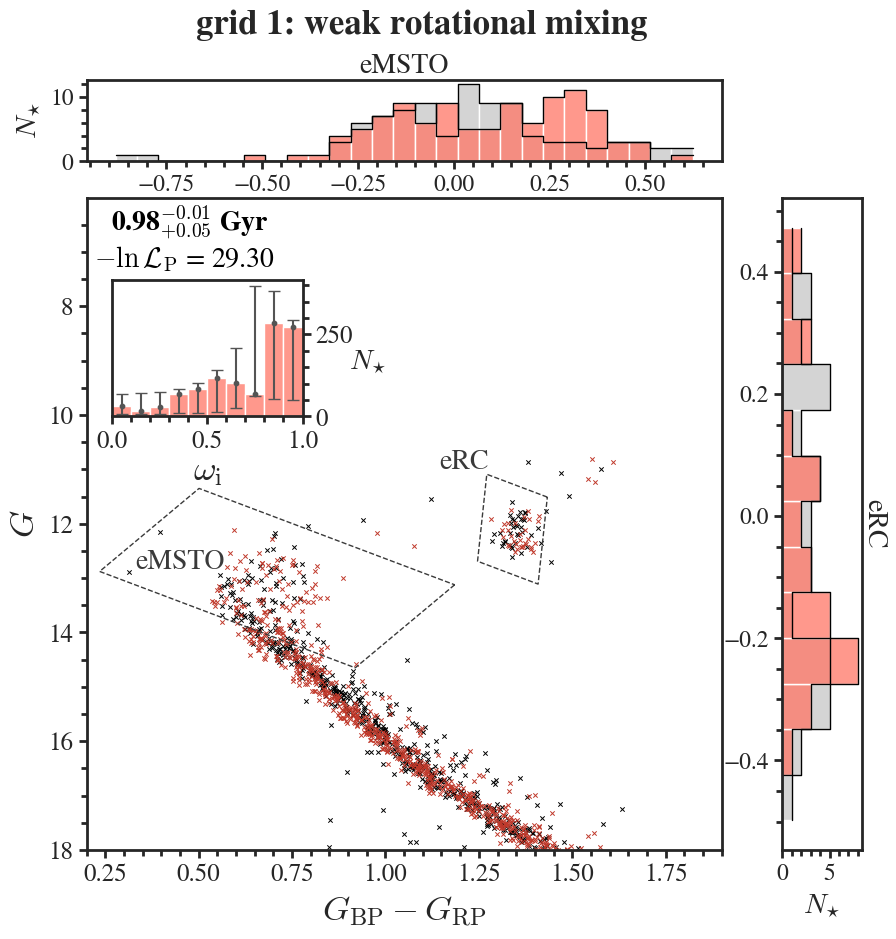}\label{fig:best_fit_NGC1817_grid1}} 
    \subfigure[]{\includegraphics[width=\columnwidth]{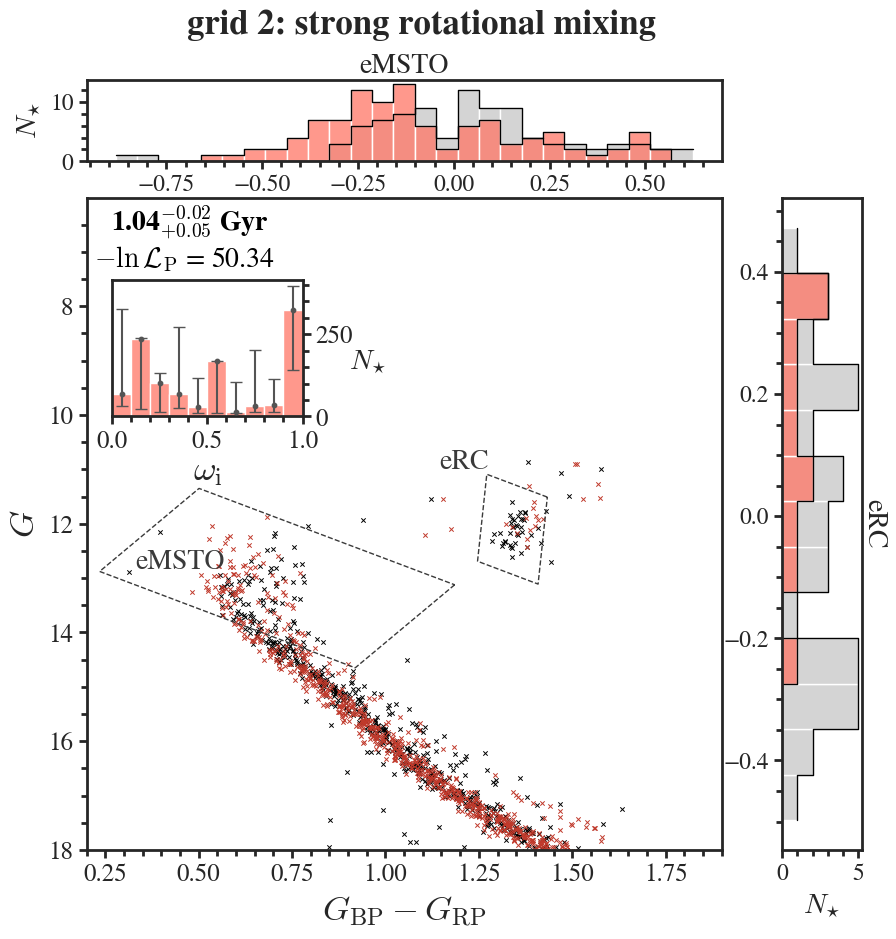}\label{fig:best_fit_NGC1817_grid2}}
    \caption{
    Left: Best fit CMD (red) obtained combining BSPs from \texttt{grid 1} stellar models (weak rotational mixing) and the observed CMD of NGC~1817 (black). Right: Best fit CMD (red) obtained combining BSPs from \texttt{grid 2} stellar models (strong rotational mixing) and observed CMD of NGC~1817 (black). The figure structure is identical to that of Fig. \ref{fig:best_fits_NGC419}.
    }
    \label{fig:best_fits_NGC1817}
\end{figure*}

By simultaneously fitting the distributions of stars in the eMSTO and eRC (Section~\ref{sec:iso_clust_construction}), for \texttt{grid 1} we found a best-fitting age of $0.98^{-0.01}_{+0.05}$~Gyr, characterized by a dominant fraction of fast rotators ($\omega_{i}\gtrsim0.85$) and a smaller contribution from slower rotators (see inset in Fig.~\ref{fig:best_fit_NGC1817_grid1}).
For \texttt{grid 2} we found a best-fitting age of $1.04^{-0.02}_{+0.05}$~Gyr, a high fraction of fast rotators with $\omega_{i}\sim0.95$ and a moderate contribution of slower rotators (see inset in Fig. \ref{fig:best_fit_NGC1817_grid2}).
The uncertainties were computed following the same procedure described in Section \ref{subsec:NGC419}.
The best-fit synthetic cluster found using \texttt{grid 1} (weak rotational mixing) gives a $-\ln \mathcal{L}_{\rm P} =29.30$, which is smaller than that found for \texttt{grid 2} (strong rotational mixing case) that gives a $-\ln \mathcal{L}_{\rm P}=50.34$.

In the case of NGC~1817, the much smaller number of stars limits the statistical significance, and caution is required. However, for the eMSTO and eRC, where the number of stars is sufficient to allow some general conclusions, the results appear similar to those obtained for NGC~419. The best-fit synthetic cluster based on \texttt{grid 1} provides a good match to the eMSTO and eRC magnitudes, colors, and widths. The best-fit synthetic cluster obtained with \texttt{grid 2} yields a clearly worse fit, showing the same issues seen for NGC~419: a significant overabundance of stars on the bright blue side of the eMSTO, a lack of stars on the faint red side, and an eRC that is on average brighter than observed. No conclusions can be drawn for the SGB and RGB, as these phases are too sparsely populated in this cluster.

Despite the limitations posed by the small number statistics in this cluster, the results still allow us to rule out strong rotational mixing, as implemented in the \texttt{grid 2} models. Not only does it produce the same type of discrepancies between the synthetic clusters and the observations as in the case of NGC419, but these discrepancies are also consistent with the expected effects of rotational mixing discussed in Section \ref{sec:rot_mix_eMSTOs_eRCs} and illustrated in Fig.~\ref{fig:base_synth_pops}.

\subsubsection{Mass-luminosity diagram of red clump stars}\label{subsubsec:NGC1817_mass_lum}
For NGC~1817, we also take advantage of the available asteroseismic parameter $\nu_{\rm{max}}$, derived from \emph{K2} photometry for 29 giants by \citet[][]{Sandquist+2020}, in combination with \emph{Gaia}~DR3 colours, magnitudes, and parallaxes, to test the predictions for the RC stellar mass range and luminosity in synthetic clusters generated from rotating stellar models.

\citet{Sandquist+2020} derived the masses and radii of these giants using the global asteroseismic parameters $\nu_{\rm{max}}$ and $\Delta\nu$, combined with photometric $T_{\rm{eff}}$ values inferred from $(b-y)$ colours, applying two standard scaling relations. However, due to the large uncertainties associated with $\nu_{\rm{max}}$ and $\Delta\nu$ measured from \emph{K2} photometry, the resulting uncertainties on stellar mass were too large to provide useful constraints. 

\begin{figure*}
    \centering
    \includegraphics[width=0.95\textwidth]{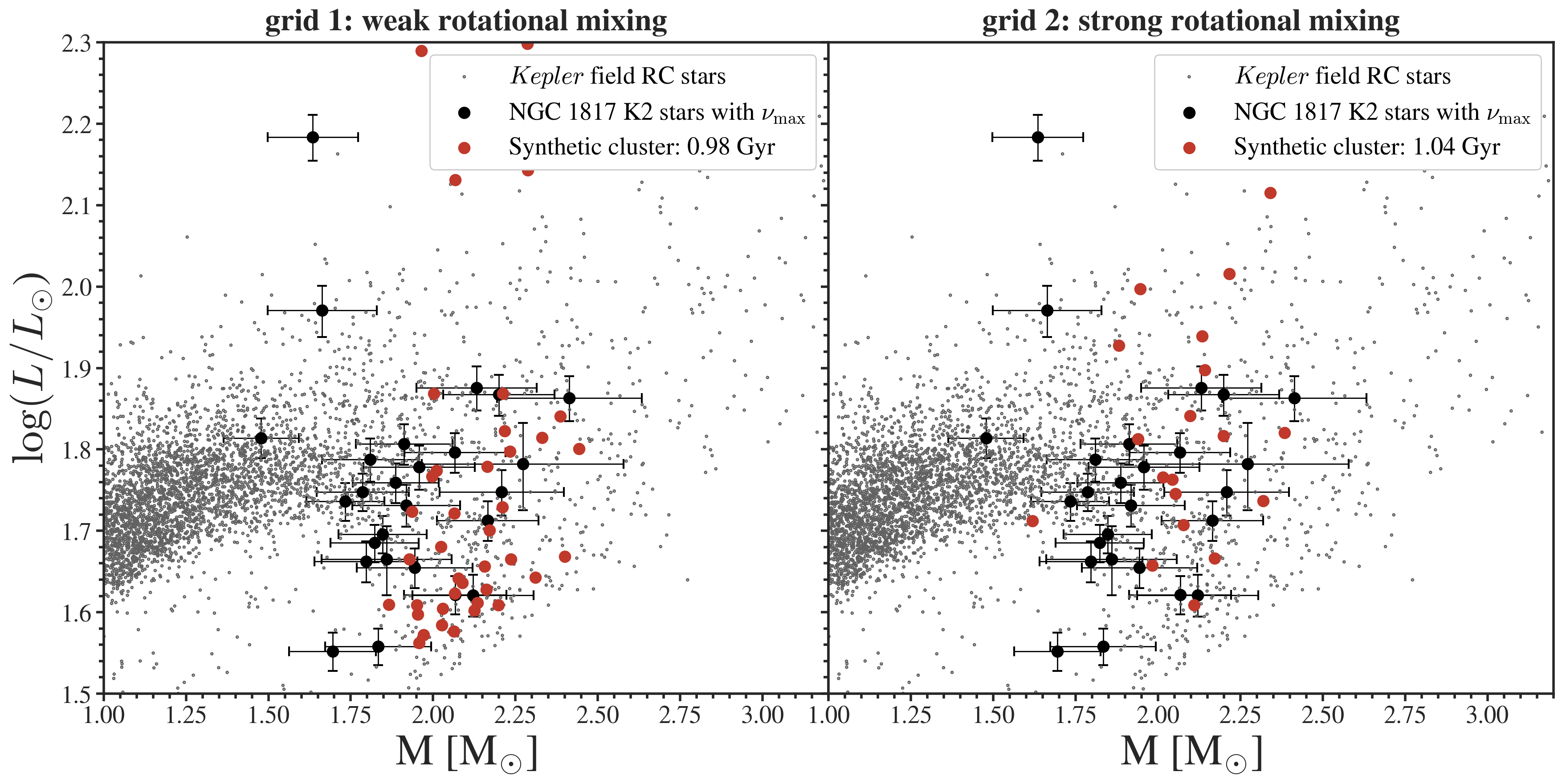}\label{fig:NGC1817_mass_lum}
    \caption{Mass–luminosity diagram of NGC~1817 giants with measured $\nu_{\rm max}$ from \emph{K2} data \citep[][]{Sandquist+2020} (black dots) and \emph{Kepler} field RC stars from the catalogue of \citet{Willett+2025submitted} (gray circles), shown in both panels. Red dots indicate the best-fit synthetic cluster stars with $\nu_{\rm max}<300$~$\mu$Hz, from \texttt{grid 1} in the left panel and from \texttt{grid 2} in the right panel.
    }
    \label{fig:mass_lum_NGC1817}
\end{figure*}

To overcome this, we derived stellar masses using the following asteroseismic scaling relation:
\begin{equation}
    \frac{M_{\rm{seis}}}{M_{\odot}} \simeq \left(\frac{\nu_{\rm{max}}}{\nu_{\rm{max,\odot}}}\right) \left(\frac{L}{L_{\odot}}\right) \left(\frac{T_{\rm{eff}}}{T_{\rm{eff,\odot}}}\right)^{-7/2},  
    \label{eq:scaling_relation}
\end{equation}  
\citep[e.g.][]{Miglio+2012}, where luminosities and effective temperatures were computed from \emph{Gaia}~DR3 colours, magnitudes, and parallaxes corrected for the zero point offset following \citet{Lindegren+2021}. This relation, which uses only $\nu_{\rm{max}}$ as the asteroseismic input and relies on precise luminosities derived from \emph{Gaia}~DR3 photometry and parallaxes, allowed us to significantly reduce the uncertainties on the resulting asteroseismic masses (see Figure~\ref{fig:mass_lum_NGC1817}).

Because the scaling relation in Equation~\ref{eq:scaling_relation} depends only on global stellar properties, it avoids potential complications related to variations in helium core masses induced by different levels of rotational mixing. Knowledge of the intrinsic mass and luminosity of RC stars provides an additional constraint on internal mixing processes in earlier evolutionary phases, particularly during the main sequence.

The synthetic cluster masses, luminosities, and effective temperatures were derived in a manner consistent with the NGC~1817 data, starting from \emph{Gaia}~DR3 colours, magnitudes, and $\nu_{\rm max}$, with the input parameters perturbed according to the observational uncertainties before applying Equation~\ref{eq:scaling_relation}.

The stellar masses and luminosities in the eRC of the best-fit synthetic cluster to NGC1817 computed with \texttt{grid 1} are consistent with the values obtained from observations (left panel of Fig.\ref{fig:mass_lum_NGC1817}), highlighting the robustness of the model. This agreement further supports the good match observed in the CMD. In particular, the distribution of masses and luminosities closely follows that of \emph{Kepler} field RC stars (gray dots in Fig. \ref{fig:mass_lum_NGC1817}) around $\sim 2.0$ \Msun{}, the mass corresponding to the minimum luminosity of He-burning stars and the transition between degenerate and non-degenerate He ignition \citep[][]{Girardi1999}. By contrast, the masses derived for NGC~1817 (using $\nu_{\rm max}$ measurements from \emph{K2} data) appear to be systematically shifted to slightly lower values compared to the bulk of the \emph{Kepler} field RC stars distribution around the transition mass. 
For the best-fit synthetic cluster based on \texttt{grid 2}, the mass range of RC stars is also consistent with that obtained from observations, while the luminosity distribution is only partly consistent and extend to higher values than observed. However, the combination of small-number statistics and mass uncertainties does not allow us to significantly prefer one model over the other in the mass-luminosity diagram.


\subsubsection{Rotation across the eMSTO of NGC~1817}\label{subsubsect:NGC1817_rotation}
\begin{figure}
    \centering
	\includegraphics[width=\columnwidth]{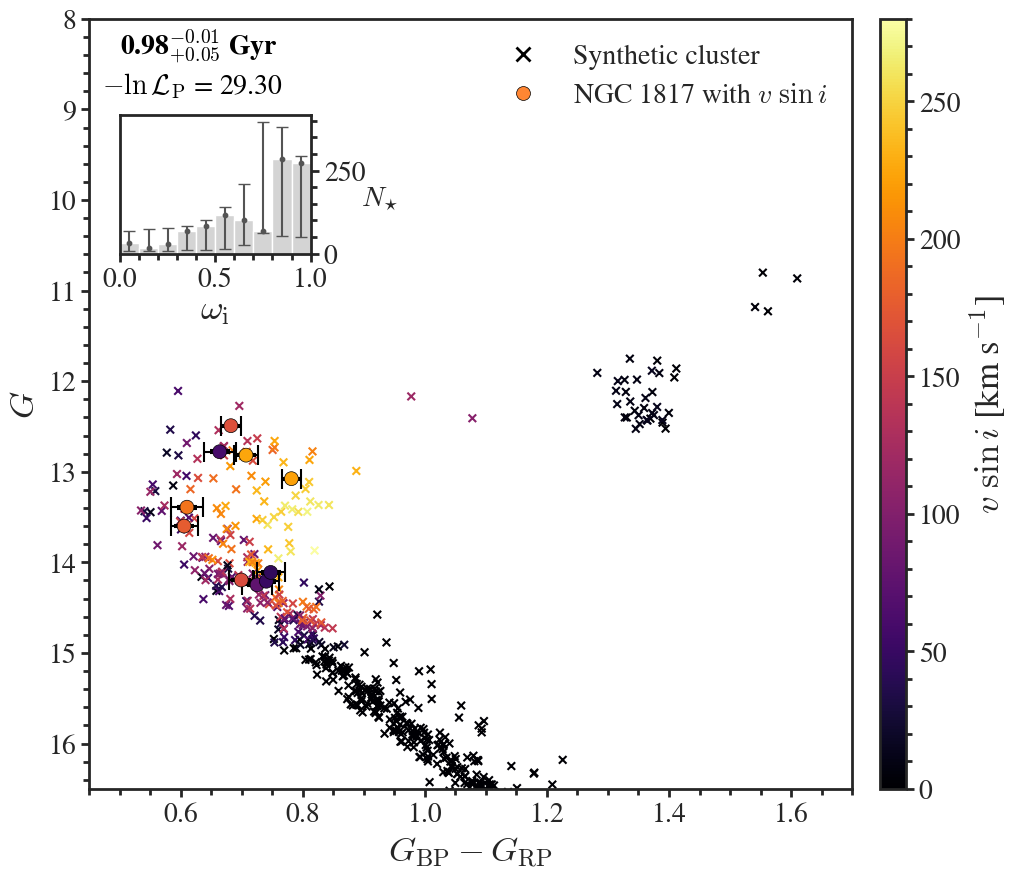}
    \caption{CMD of our best-fit synthetic cluster for NGC~1817 from \texttt{grid 1} stellar models (left panel of Fig.~\ref{fig:best_fits_NGC1817}). Synthetic stars are shown as crosses and coloured by their $v \sin i$ values. Overlaid are the 10 stars with $v \sin i$ measurements from \citet{Molenda-Zakowicz+2009} (filled circles), using the same colour scale. The inset shows the histogram of initial rotation rates for the stars in the best-fit synthetic cluster.}

    \label{fig:NGC1817_vsini}
\end{figure}

\citet{Molenda-Zakowicz+2009} measured $v\sin i$ values ranging from 45 to 225km~s$^{-1}$ for 11 $\delta$Sct variables located across the MSTO of NGC~1817. We cross-matched these stars with our \emph{Gaia}~DR3 catalogue of cluster members using \textsc{Topcat} \citep[][]{Taylor2005}, finding 10 matches. All of these, except for the star labelled V12 in \citet{Molenda-Zakowicz+2009}, were included in our list of cluster members.
These 10 $\delta$~Sct variables are shown in Figure~\ref{fig:NGC1817_vsini}, plotted on top of the our best-fit synthetic cluster and colour-coded by their $v\sin i$ values using the same colour scale as for the synthetic cluster.
Although the sample size is small and does not allow for statistically significant conclusions, the observed $v \sin i$ distribution across the eMSTO appears consistent with the scenario seen in NGC~419, where stars on the blue side have lower $v \sin i$ and stars on the red side higher values, including the fact that the largest $v \sin i$ values predicted by the synthetic cluster for the red side tend to exceed the measured values in the same CMD region. Similar considerations as those discussed for the comparison between measured and synthetic $v \sin i$ in NGC~419 also apply here.
% End of The impact of rotational mixing on extended main sequence turn-offs and extended red clumps

\section{Conclusions}\label{sec:conclusions}
We have constructed two homogeneous grids of rotating stellar models with high resolution in both initial mass and rotation rate (see Section \ref{sec:models_grids}). These grids were computed using identical input physics, differing only in the treatment of convective core overshooting and the implementation of rotational mixing. One grid adopts the standard \texttt{MESA} prescription for rotation-induced mixing (\verb|grid 1|), while the other is calibrated to reproduce the efficiency of rotational mixing in the \texttt{GENEC} models of \citet{Georgy+2013} (\verb|grid 2|). Both prescriptions are widely used in stellar models grids applied to studies of eMSTOs and stellar rotation. Stellar models from \verb|grid 1| exhibit very limited effects of rotational mixing compared to those from \verb|grid 2| (see Section \ref{subsubsec:rot_mix} and Figure \ref{fig:grid1_vs_grid2}).

We used these grids to generate a set of statistically independent single-age synthetic stellar populations, referred to as "Base Stellar Populations", covering the full range of initial rotation rates from non-rotating to
near-critical.

By analyzing the CMDs and $v \sin i$ distributions of these BSPs across different ages, rotation rates, and convective core overshooting efficiencies, we assessed the relative roles of rotational mixing, gravity darkening, overshooting, and age in shaping eMSTOs and eRCs (see Section \ref{sec:rot_mix_eMSTOs_eRCs} and Figure \ref{fig:base_synth_pops}). This framework provides a natural explanation for the presence of both eMSTOs and eRCs in single-age populations, as well as for clusters that display eMSTOs but compact RCs.

We used an optimization algorithm to determine, the best combination of BSPs to fit the CMDs of two star clusters: the Small Magellanic Cloud cluster NGC 419 and the Milky Way open cluster NGC 1817, both of which display prominent eMSTOs and eRCs, and are known to host fast-rotating stars on the eMSTO (see Sections~\ref{subsec:NGC419} and ~\ref{subsec:NGC1817}).
For NGC~1817, we also obtained a precise estimate of the color excess $E(B-V)$ from five red giant stars, using spectroscopic $T_{\rm{eff}}$ and $\log g$ from APOGEE~DR17, combined with \emph{Gaia}~DR3 $G_{\rm{BP}}-G_{\rm{RP}}$ color index. We further derived the most precise estimates to date of the RC star masses in this cluster, by combining $\nu_{\rm{max}}$ measurements from \citet{Sandquist+2020} with luminosities and $T_{\rm{eff}}$ values derived from \emph{Gaia}~DR3 photometry and parallaxes.
Our analysis of NGC~419 and NGC~1817 shows that the best-fit synthetic clusters are obtained with models that include weak rotational mixing rather than strong mixing (see Figures \ref{fig:best_fits_NGC419} and \ref{fig:best_fits_NGC1817}), and that they require a high fraction of fast rotators with $\omega_{i} \gtrsim 0.85$, close to critical rotation. A similarly high fraction of fast rotators in NGC~419 and in other intermediate-age clusters has also been found by \citet{Ettorre+2025}. 

NGC~419 and NGC~1817, while similar in age and both exhibiting eMSTO and eRC features, differ in important respects: they reside in distinct environments, the SMC and the Milky Way respectively, have different metallicities, and were observed independently with different instruments and processed through separate data reduction pipelines. Despite these differences, both clusters consistently indicate that strong rotational mixing, which would significantly modify the envelope composition, is disfavored because it produces changes in luminosity and temperature that lead to worse CMD fits. Our results are further supported by the mass and luminosity range of RC stars derived for NGC~1817 (Figure \ref{fig:mass_lum_NGC1817}). It is not excluded, however, that rotational mixing could slightly increase the He-core mass at hydrogen exhaustion, potentially enough to affect the transition from degenerate to non-degenerate helium ignition (see Section \ref{sec:rot_mix_eMSTOs_eRCs} and Figure \ref{fig:base_synth_pops}).

A similar result was found by \citet{Brogaard+2023} for the core He-burning stars in the open cluster NGC~6866, who concluded that the actual effects of rotation are smaller than those predicted by 1D \verb|GENEC| models, which overestimate stellar radii compared to values derived from asteroseismology.
It remains to be verified whether the same scenario is consistently supported by surface chemical abundance measurements of stars in these clusters. 

The best-fit synthetic clusters with weak rotational mixing are broadly consistent with the observations, but certain aspects, including systematically higher $v \sin i$ values on the red side of the eMSTOs, show discrepancies (Sections \ref{subsubsect:NGC419_rotation}, \ref{subsubsect:NGC1817_rotation}). Similar discrepancies have been identified in other studies \citep{Bastian+2018, Kamann+2020, Kamann+2023}. In some cases, even stars rotating near the critical limit fail to reproduce the full observed width of the eMSTO, suggesting that additional mechanisms may be required \citep{Goudfrooij+2017, Gossage+2019, Lipatov+2022}. The excess of stars on the blue-bright side of the best-fit CMD (Fig.~\ref{fig:best_fit_NGC419_grid1}) strengthen the argument that the predicted rotation rates may be artificially too high due to missing or poorly modelled aspects of stellar rotation. We proposed a possible connection with the observed Be \citep[][]{Milone+2018, Bastian+2016} and UV-dim stars phenomenon \citep[][]{Martocchia+2023, Milone+2023b}: a large fraction of fast rotators may develop decretion discs at high rotation rates, with dust grains forming in the outer regions \citep[][]{D'Antona+2023, He+2025} and absorbing light from stars seen close to equator-on, shifting their positions on the eMSTO. To explore this, we performed a schematic test in which we increased the reddening by $\Delta E(B-V)\sim0.1$ mag for stars with $v/v_{\mathrm{crit}}\gtrsim0.7$ observed within $\sim15^{\circ}$ of the equatorial plane. This broadened the eMSTO while preserving realistic $v \sin i$ values and reducing the excess of blue-bright stars. While highly simplified, this experiment suggests that disc absorption could influence inferred rotation distributions. Testing this possibility will require both detailed disc modelling and more complete star-by-star spectroscopic and photometric datasets, including $v\sin i$ measurements across eMSTOs.
% End of Conclusions

\section*{Acknowledgements}
This project was undertaken with the assistance of resources and services from the National Computational Infrastructure (NCI), which is supported by the Australian Government. Access to NCI was supported by the University of Newcastle.
HS is the recipient of an Australian Research Council Future Fellowship Award (project number FT220100330). This research is funded by this grant from the Australian Government.
HS \& LM acknowledge the Awabakal people, the traditional custodians of the unceded land on which their research was undertaken. AM acknowledges support from the ERC Consolidator Grant funding scheme (project ASTEROCHRONOMETRY, G.A. n. 772293 \url{http://www.asterochronometry.eu}). G.B. acknowledges fundings from the Fonds National de la Recherche Scientifique (FNRS) as a postdoctoral researcher. PE acknowledges support from the SNF grant No 219745.

We thank the anonymous referee for the suggestions that led to a significant improvement of the quality of this work.
We sincerely thank Léo Girardi for generously providing the reduced data for NGC~419 published in \citet{Girardi+2009}.
LM thanks Diego Bossini for his help with bolometric corrections, Marco Tailo, Lorenzo Briganti, and Alessandro Mazzi for valuable discussions and suggestions. LM also gratefully acknowledges Georges Meynet, the STAREX group members, and the Observatoire de Genève for their warm hospitality during a research stay, where the idea for this work first took shape.
%%%%%%%%%%%%%%%%%%%%%%%%%%%%%%%%%%%%%%%%%%%%%%%%%%
\section*{Data Availability}

The \texttt{MESA} release \texttt{r-24.03.1} used in this work is available at \url{https://doi.org/10.5281/zenodo.10783349}.
The \texttt{MESA} SDK required to run the code is available at: \url{https://doi.org/10.5281/zenodo.2603175}.
We have made the inlists and \texttt{run\textunderscore star\textunderscore extras.f90} used to generate the stellar models, publicly available on Zenodo: \url{https://doi.org/10.5281/zenodo.15601753}. The digital data underlying this article are available from the corresponding author upon reasonable request.

This work made use of the following software:
Python (\url{https://www.python.org/}), NumPy \citep[][]{harris+2020}, SciPy \citep[][]{SciPy}, Pandas \citep[][]{Pandas, mckinney2010}, Matplotlib \citep[][]{Hunter2007}, \texttt{PyGAD} \citep[][]{PyGAD}, \texttt{MESA} \citep[][]{Paxton+2011, Paxton+2013, Paxton+2015, Paxton+2018, Paxton+2019, Jermyn+2023}, \texttt{SYCLIST} \citep[][]{Georgy+2014}, and \textsc{Topcat} \citep[][]{Taylor2005}.



\bibliographystyle{mnras}
\bibliography{bibliography} % if your bibtex file is called example.bib

%\begin{thebibliography}{99}
%\end{thebibliography}

%%%%%%%%%%%%%%%%%%%%%%%%%%%%%%%%%%%%%%%%%%%%%%%%%%

%%%%%%%%%%%%%%%%% APPENDICES %%%%%%%%%%%%%%%%%%%%%

\appendix
\section{Calibration of the diffusion coefficients to mimic rotational mixing in \texttt{GENEC} models}\label{app:rot_mix_calib}

In this section, we describe the method used to calibrate the fully diffusive custom rotational mixing prescription in \verb|MESA| active during the MS phase implemented in \verb|grid 2| models. This custom prescription is designed to mimic rotational mixing in \verb|GENEC| models from \citet{Georgy+2013} (hereafter referred to as the reference grid). The evolution of angular momentum and chemical element profiles, due to rotation-driven instabilities in \verb|MESA|, is governed by two diffusion equations, which are computationally efficient to solve:  
\begin{equation}
\frac{\partial \Omega}{\partial t} = \frac{1}{\rho r^4} \frac{\partial}{\partial r} \left[ \rho r^4 \left(D_{\rm{mix}}/f_c\right) \frac{\partial \Omega}{\partial r} \right],
\label{am_transport_MESA}
\end{equation}

\begin{equation}
\frac{\partial X_i}{\partial t} = \frac{1}{\rho r^2} \frac{\partial}{\partial r} \left( \rho r^2 D_{\rm{mix}} \frac{\partial X_i}{\partial r} \right),
\label{chem_transport_MESA}
\end{equation}
where $X_{i}$ is the mass fraction of nuclide $i$, $D_{\rm{mix}}$ is the total diffusion coefficient (a sum of all diffusion coefficients for the transport processes considered), and $f_c$ is the ratio of the diffusion coefficient to the turbulent viscosity, defined as $f_c = D_{\rm{mix}} / \nu$. This ratio is calibrated using observational constraints. For details on the calculation of the diffusion coefficients associated with the transport processes in this diffusive formalism, see \citet{Heger+2000}.

The advective-diffusive formalism for rotation-induced mixing in \verb|GENEC| \citep[][]{Eggenberger+2008} describes angular momentum transport with an advective-diffusive equation:  
\begin{equation}
\rho \frac{d}{dt} \left[ r^2 \Omega \right] = \frac{1}{5 r^2} \frac{\partial}{\partial r} \left[ \rho r^4 \Omega U \right] + \frac{1}{r^4} \frac{\partial}{\partial r} \left[ \rho D_{\rm{shear}} r^4 \frac{\partial \Omega}{\partial r} \right],
\label{am_transport_GENEC}
\end{equation}
and chemical element transport with a diffusive equation:
\begin{equation}
\rho \frac{\partial X_i}{\partial t} = \frac{1}{r^2} \frac{\partial}{\partial r} \left[ r^2 \rho \left( D_{\rm{eff}} + D_{\rm{shear}} \right) \frac{\partial X_i}{\partial r} \right] = \frac{1}{r^2} \frac{\partial}{\partial r} \left[ r^2 \rho D_{\rm{mix}} \frac{\partial X_i}{\partial r} \right].
\label{chem_transport_GENEC}
\end{equation}
Here, $U$ is the radial component of the meridional circulation velocity, $D_{\rm{eff}}$ is the effective diffusion coefficient for the transport of chemical elements due to the combined effects of meridional circulation and horizontal turbulence, and $D_{\rm{shear}}$ is the diffusion coefficient for the radial shear instability \citep[for details see][]{Eggenberger+2008}. 
The diffusion coefficients in the \verb|GENEC| formalism are derived using self-consistent physical models based on reasonable assumptions \citep{Maeder2009,Nandal+2024}. These coefficients are complicated functions, depending on the local physical conditions within the star and the velocity of meridional circulation  \citep[see][for details]{Chaboyer&Zahn1992,Meynet&Maeder1997,Maeder2009}.  

\begin{figure*}
	\includegraphics[width=0.9\textwidth]{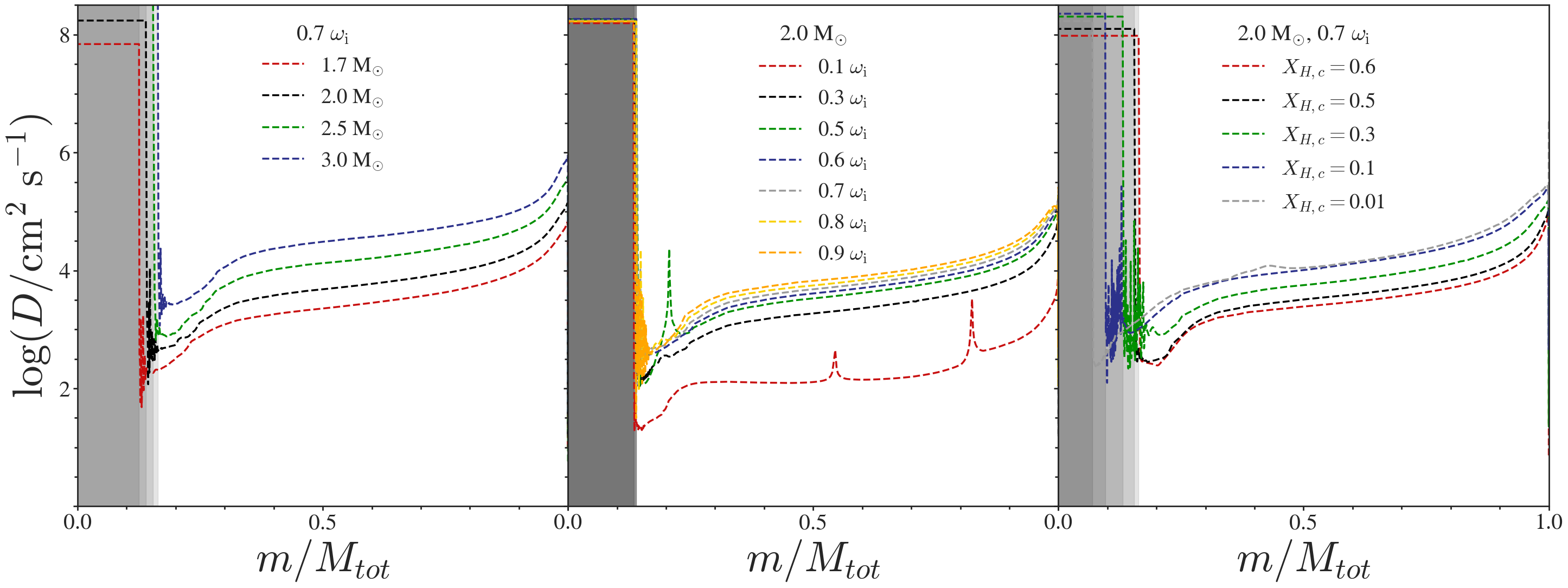}
    \caption{Left: $D_{\rm{mix}}$ profiles for \texttt{GENEC} models with masses of 1.7, 2.0, 2.5, and 3.0 \Msun{}, each with an initial rotation rate of 0.7 $\omega$, evaluated at approximately halfway through the MS phase ($X_{\rm{^1H}, c} = 0.35$). Center: $D_{\rm{mix}}$ profiles for a 2.0 \Msun{} \texttt{GENEC} model with initial rotation rates of 0.1, 0.3, 0.5, 0.6, 0.7, 0.8, and 0.9 $\omega$, evaluated at approximately halfway through the MS phase ($X_{\rm{^1H}, c} = 0.35$). Right: $D_{\rm{mix}}$ profiles for a 2.0 \Msun{} \texttt{GENEC} model with an initial rotation rate of 0.7 $\omega$, evaluated at different evolutionary stages during the MS, corresponding to central hydrogen mass fractions of $X_{\rm{^1H}, c} = 0.6, 0.5, 0.3, 0.1, 0.01$.}
    \label{fig:D_mix_change}
\end{figure*}

\begin{figure}
    \centering
	\includegraphics[width=0.7\columnwidth]{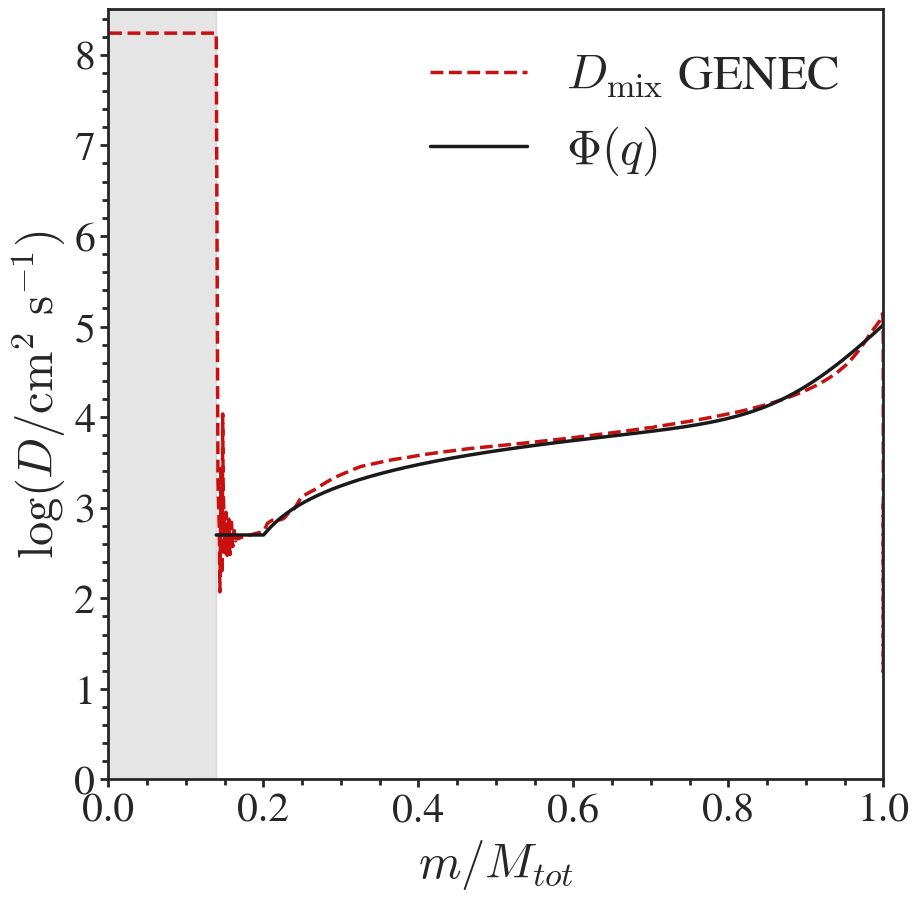}
    \caption{Best fit of the parametric function (Eq. \ref{f(q)}) to the $D_{\rm{mix}}$ profile of the 2.0 \Msun{} model, selected as the representative model to determine the fixed shape parameters $a$, $b$, $c$, $d$, $e$, and $f$.}
    \label{fig:D_mix_fit}
\end{figure}

An inspection of the diffusion coefficients $D_{\rm{mix}}$ (see Fig.~\ref{fig:D_mix_change}) across the range of masses and rotation rates covered by the reference grid, reveals that the qualitative shape of the diffusion coefficient profiles remains consistent despite the presence of some short-lived spikes. The most notable variation is that $D_{\rm{mix}}$ scales approximately by an overall factor that increases with stellar mass, initial rotation rate $\omega_{\rm{i}}$, and progression along the MS phase (see Figure \ref{fig:D_mix_change}).
The goal of this custom rotational mixing prescription is not to exactly replicate the advective-diffusive implementation of \verb|GENEC| models but rather to develop a computationally efficient approximation that qualitatively reproduces the main properties of \verb|GENEC| rotating models relative to their non-rotating counterparts. Specifically, the aim is to reproduce the increase in MS lifetime and changes to the surface chemical element within a small tolerance.  
We consider the $D_{\rm{mix}}$ profile to consist of two distinct zones: the near-core zone, dominated by $D_{\rm{eff}}$, which accounts for the combined effects of meridional circulation and horizontal turbulence, and the envelope zone, dominated by $D_{\rm{shear}}$.  
The value of $D_{\rm{mix}}$ near the core primarily influences the MS lifetime and the increase in MS luminosity, whereas the value in the envelope  mainly affects surface chemical element abundance variations.  
To approximate the total diffusion coefficient $D_{\rm{mix}}$ profile in the radiative zone, we adopt the following parametric function:  
\begin{equation}
\Phi(q) = 
\begin{cases} 
\sigma \cdot \left[ a \cdot q - (q_{\rm{cc}} + b) + 10^{c \cdot q - d} + 10^{e \cdot q - f} \right] & \text{if } \Phi > \beta, \\
\beta & \text{if } \Phi \leq \beta,
\end{cases}
\label{f(q)}
\end{equation}
where $q = m / M$, with $m$ representing the mass within a given radius, $M$ the total stellar mass and $q_{\rm{cc}}$ the mass coordinate of the convective core boundary, including the overshooting region.  
The parameters $a$, $b$, $c$, $d$, $e$, and $f$ determine the shape of the function and are fixed via least-squares minimization to fit the  $D_{\rm{mix}}$ profile of a representative model\footnote{For calibration, we use a 2.0 \Msun{} model initialized with 0.7 $\omega_{\rm{i}}$ at the midpoint of the MS phase.} (see Fig. ~\ref{fig:D_mix_fit}).
The fitted parameter values are listed in Table \ref{tab:mathssymbols}.  
\begin{table}
\caption{Fixed parameters of the parametric function (Equation \ref{f(q)}) obtained through least squares minimization to fit the $D_{\rm{mix}}$ profile of a representative 2.0 \Msun{} model with an initial rotation rate of 0.7 $\omega$.}
\label{tab:mathssymbols}
\begin{tabular}{ll}
\hline
Fixed parameter & Value \\
\hline
$a$ & 1.2480576095175386 \\
$b$ & 0.0615015714089750 \\
$c$ & 8.6996929817348025 \\
$d$ & 8.0272056977252149 \\
$e$ & 8.6995994374779517 \\
$f$ & 8.0277869134778879 \\
\hline
\end{tabular}
\end{table}  
We discuss the calibration constants, $\sigma$ and $\beta$, further in this Appendix.
%The near-core zone is calibrated (or scaled) with $\beta$, while the envelope zone is calibrated by $\sigma$. 

\begin{figure}
    \centering
	\includegraphics[width=0.7\columnwidth]{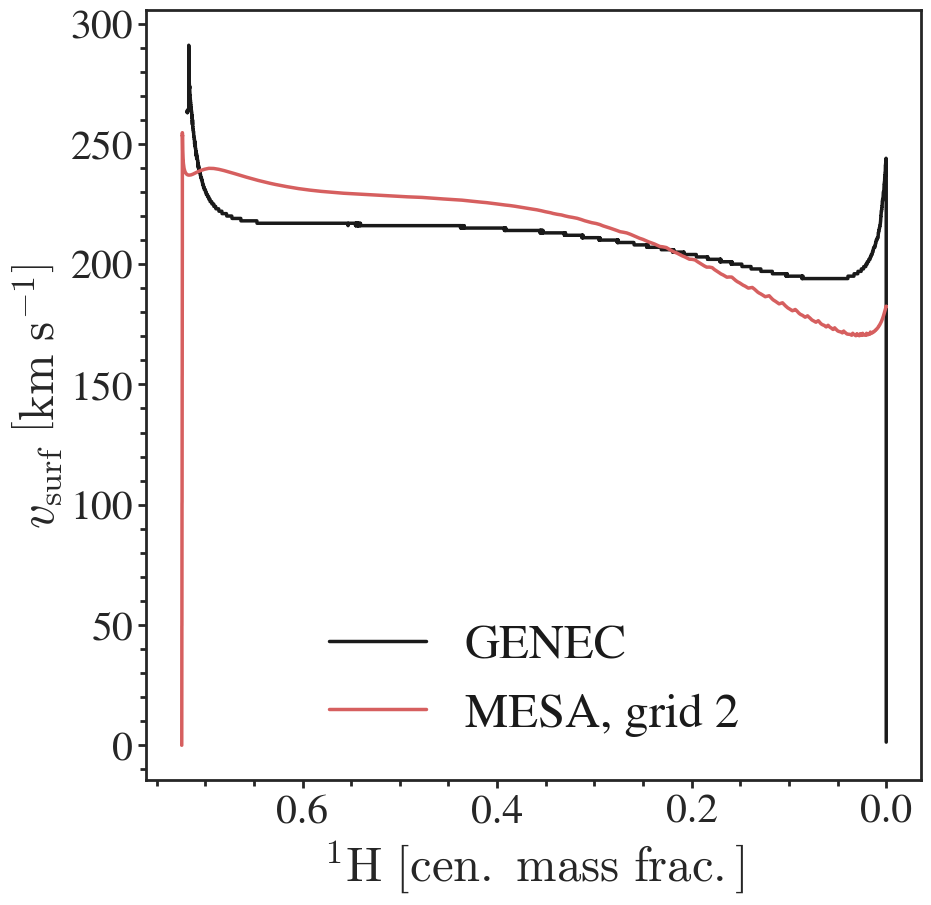}
    \caption{Comparison of the evolution of the equatorial velocity between a \texttt{GENEC} reference model (black) of 2.5 \Msun{}, rotating with 0.8 $\omega_{\rm{i}}$, and a \texttt{MESA} model of the same mass and rotation rate from \texttt{grid 2} (red), calibrated using the technique described in Appendix \ref{app:rot_mix_calib}.}
    \label{fig:v_eq_GENEC_MESA}
\end{figure}

In the \verb'MESA' models of \verb'grid 2', we define the total diffusion coefficient for chemical transport as:  
\begin{equation}
D_{\rm{mix}}(q) = \chi_{H,c} \cdot \Phi(q),
\end{equation}  
and for the transport of angular momentum, we impose the angular momentum diffusion coefficient $\nu_{\rm{AM}}$ as: 
\begin{equation}
\nu_{\rm{AM}}(q) = 5 \cdot \chi_{H,c} \cdot \Phi(q),
\end{equation}  
where $\chi_{H,c}$ is a factor that linearly increases from 0.01 to 1 over the MS phase, depending on the central hydrogen mass fraction. This approach qualitatively reproduces the overall increase in $D_{\rm{mix}}$ during the MS phase, consistent with trends observed in the \verb'GENEC' models, as shown in the right panel of Figure \ref{fig:D_mix_change}. 
The scaling factor of 5 applied to the angular momentum diffusion coefficient ensures a qualitatively acceptable match to the evolution of surface equatorial velocity across all \verb'GENEC' models in the reference grid (Fig.~\ref{fig:v_eq_GENEC_MESA}). 
%An example of the comparison between the equatorial velocity in a calibrated \verb|MESA| model and that in a reference \verb|GENEC| model is illustrated in Figure~\ref{fig:v_eq_GENEC_MESA}.  

Using a good approximation of \verb|GENEC| $D_{\rm{mix}}$ values in \verb|MESA| does not guarantee similar increases in MS lifetime or surface chemical element variations. This discrepancy arises from inherent differences in the codes and assumptions in their input physics. Consequently, we treat $\beta$ and $\sigma$ as free parameters to be calibrated in order to reproduce similar increases in MS lifetime and surface chemical element variations. These parameters scale the diffusion coefficients in the near-core and envelope zones, respectively, without altering their overall shape.  

We calibrated the profiles using the Nelder-Mead optimization algorithm \citep[][]{Nelder&Mead1965}, which minimizes the residuals of two \emph{calibration parameters}, $\Pi_1$ and $\Pi_2$:  
\begin{gather}
\Pi_1 = \frac{\tau^{\omega_{\rm{i}}}_{\rm{TAMS}}}{\tau^{0.0}_{\rm{TAMS}}}, \\
\Pi_2 = \frac{\left(X_{\rm{N}} / X_{\rm{H}}\right)^{\omega_{\rm{i}}}_{\rm{TAMS}}}{\left(X_{\rm{N}} / X_{\rm{H}}\right)^{0.0}_{\rm{TAMS}},}  
\end{gather}  
where $\tau^{\omega_{\rm{i}}}_{\rm{TAMS}}$ and $\left(X_{\rm{N}} / X_{\rm{H}}\right)^{\omega_{\rm{i}}}_{\rm{TAMS}}$ are the age and the surface nitrogen-to-hydrogen ratio, respectively, at the terminal age main sequence (TAMS) of the model with an initial rotation rate $\omega_{\rm{i}}$.  
The residuals to be minimized are:  
\begin{gather}
\Delta \Pi_1 = |\Pi^{\rm{MESA}}_1 - \Pi^{\rm{GENEC}}_1|, \\
\Delta \Pi_2 = |\Pi^{\rm{MESA}}_2 - \Pi^{\rm{GENEC}}_2|.  
\end{gather}  
Calibration is considered successful when $\Delta \Pi_1 < 0.005$ and $\Delta \Pi_2 < 0.01$.
\begin{figure}
    \centering
	\includegraphics[width=0.6\columnwidth]{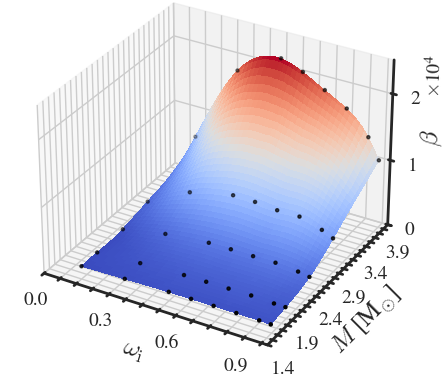}
    \includegraphics[width=0.6\columnwidth]{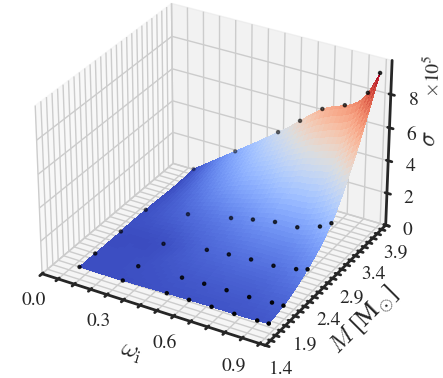}
    \caption{Top: Calibration results for the parameter $\beta$ in solar metallicity models of \texttt{grid 2}. Bottom: Calibration results for the parameter $\sigma$ in solar metallicity models of \texttt{grid 2}.
    In both panels, black dots represent the calibrated values for the reference grid models, while the values of $\beta$ and $\sigma$ at intermediate points are obtained through cubic spline interpolation. Color map reflects the value of $\beta$ and $\sigma$ respectively.}
    \label{fig:calib_params}
\end{figure}
Calibration is performed on solar metallicity ($Z = 0.014$) reference models from the \citet{Georgy+2013} grid, spanning 1.7 to 4.0 \Msun{} and $0.1 \leq \omega_{\rm{ini}} \leq 0.95$. To compute models at intermediate masses, we interpolate the calibrated diffusion coefficients using a cubic spline. Figure \ref{fig:calib_params} presents the calibrated values of $\beta$ and $\sigma$ for the reference models (black dots) along with interpolated values. For masses below the lower limit of 1.7 \Msun{} (down to 1.4 \Msun{}), we linearly extrapolate the calibrated diffusion coefficients using the slope of the cubic spline at 1.7 \Msun{}. However, due to this extrapolation, conclusions should be avoided for masses below 1.7 \Msun{}.  
For lower metallicities, we adopt the same calibrated values as at solar metallicity, since \citet{Georgy+2013} found only a very weak dependence of the diffusion coefficients on $Z$.

\begin{figure*}
    \centering
	\includegraphics[width=0.6\textwidth]{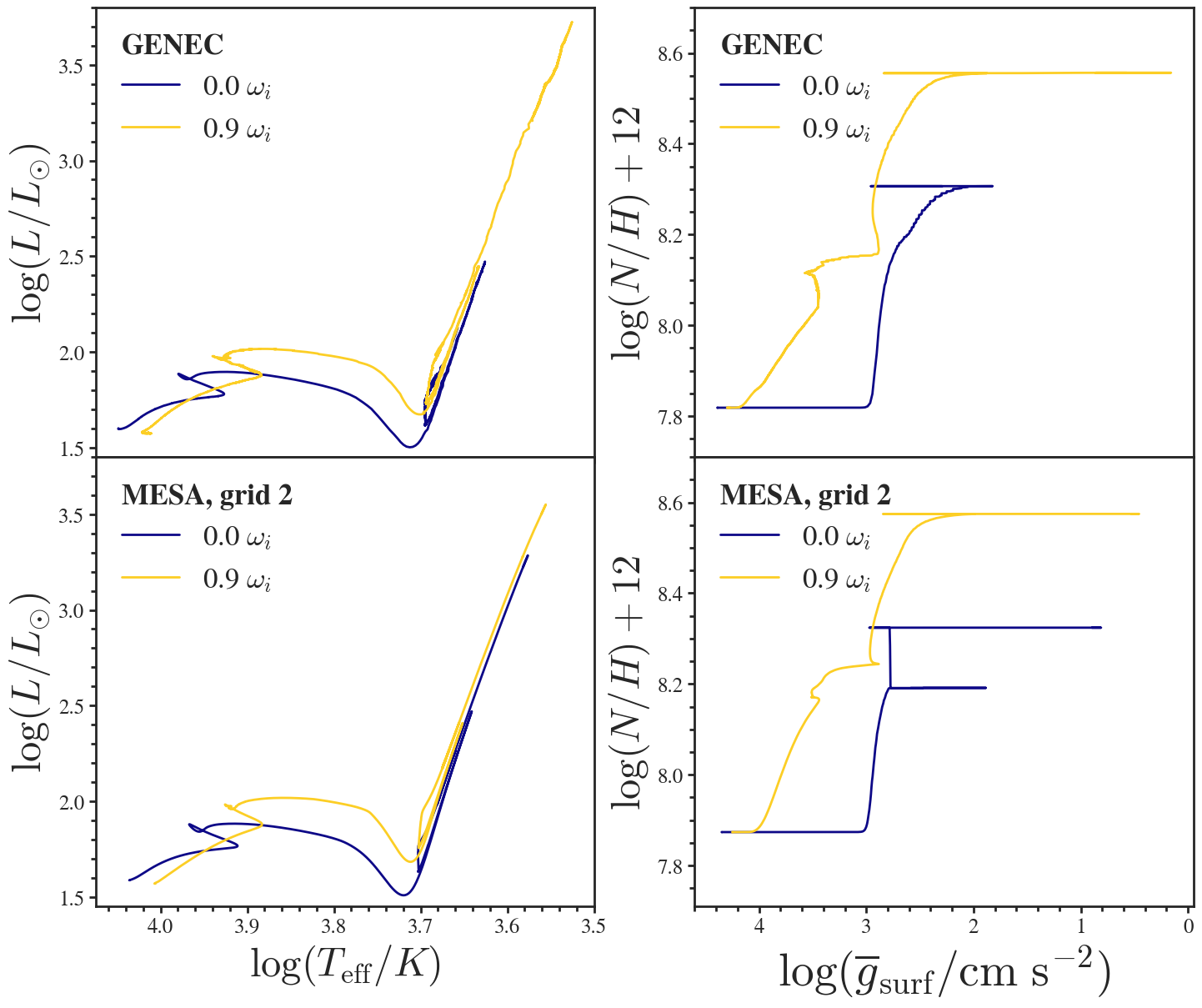}
    \caption{Left: Comparison of evolutionary tracks in the HR diagram for 2.5~\Msun{} stars from \texttt{GENEC} reference models (top) rotating at $0.9\,\omega_{\rm{i}}$ and non-rotating, and from \texttt{MESA} models (bottom) with the same mass and rotation rate, taken from \texttt{grid 2} and calibrated as described in Appendix~\ref{app:rot_mix_calib}. Right: Comparison of the evolution of the surface nitrogen-to-hydrogen abundance ratio for 2.5~\Msun{} stars from \texttt{GENEC} reference models (top) rotating at $0.8\,\omega_{\rm{i}}$ and non-rotating, and from \texttt{MESA} models (bottom) from \texttt{grid 2}, calibrated using the same method.}
    \label{fig:comparison_GENEC_MESA}
\end{figure*}

We activated this custom implementation of rotation-induced mixing during the MS phase and, at TAMS, gradually transitioned over 500 timesteps, to the default implementation. Notably, \verb|GENEC| models also transition to a fully diffusive implementation after the MS phase \citep[][]{Nandal+2024}.  
With this calibration, we achieve a good approximation of the key properties of rotating \texttt{GENEC} models using \texttt{MESA}. By construction, main-sequence lifetimes and surface chemical abundance variations are well reproduced (see right panel of Fig.\ref{fig:comparison_GENEC_MESA}). In addition, the evolution in the HR diagram and the equatorial velocity are reasonably well reproduced, as shown in the left panel of Fig.\ref{fig:comparison_GENEC_MESA} and in Fig.\ref{fig:v_eq_GENEC_MESA}. In Fig.\ref{fig:comparison_GENEC_MESA}, we compare 2.5~\Msun{} models, which are the lowest-mass \texttt{GENEC} models that include the helium-burning phase.

\section{Processing evolutionary tracks to create a uniform basis for interpolation} \label{track_sampling}

The evolutionary tracks of \verb|grid 1| and \verb|grid 2| are processed using a method closely following that described in \citet{Dotter2016}, but adapted to the \verb|SYCLIST| format. This method is based on "equivalent evolutionary points" \citep[EEPs, see][]{Dotter2016}, a series of points that can be identified across all stellar evolutionary tracks. We define eight primary EEPs as follows:

\begin{enumerate}
    \item[1.] ZAMS: The first point after the H-burning luminosity that exceeds 99.9\% of the total luminosity and before the central H mass fraction decreases by 0.0015 from its initial value.  
    \item[2.] Coolest point of the MS: The point at the lowest temperature before the central H mass fraction falls below $1.7 \cdot 10^{-4}$.
    \item[3.] Turn-off: The hottest point between point 2 and the point when the central H mass fraction falls below \(1 \cdot 10^{-10}\).
    \item[4.] RGB base: The lowest luminosity point in the post-MS phase (after point 3).
    \item[5.] RGB tip: The highest luminosity point after point 4 and before the center becomes convective due to He burning.
    \item[6.] Start of the stable Core Helium Burning: The minimum luminosity of the Core Helium Burning phase.
    \item[7.] End of Core Helium Burning: The first point after point 6 when convection due to He burning in the center stops.
    \item[8.] First thermal pulse: The end of the evolutionary track.
\end{enumerate}
Between the primary EEPs, we defined secondary EEPs (Table~\ref{tab:secondary_EEPs}) to accurately capture the morphology of each segment of the evolutionary track. The method adopted is identical to that described in Section~2.2 of \citet{Dotter2016}. At the low- and high-mass ends of the mass range covered by our grid, some of the primary EEPs can no longer be defined. In these cases, we omit the corresponding primary EEP and instead fill the interval with the same number of secondary EEPs.
\begin{table}
\caption{Number of secondary EEPs in each segment of the evolutionary track.}
\label{tab:secondary_EEPs}
\begin{tabular}{ll}
\hline
Segment & Secondary EEPs \\
\hline
1-2 & 80 \\
2-3 & 30 \\
3-4 & 50 \\
4-5 & 50 \\
5-6 & 40 \\
6-7 & 90 \\
7-8 & 60 \\
\hline
\end{tabular}
\end{table}
 
\begin{figure*}
    \centering
	\includegraphics[width=0.6\textwidth]{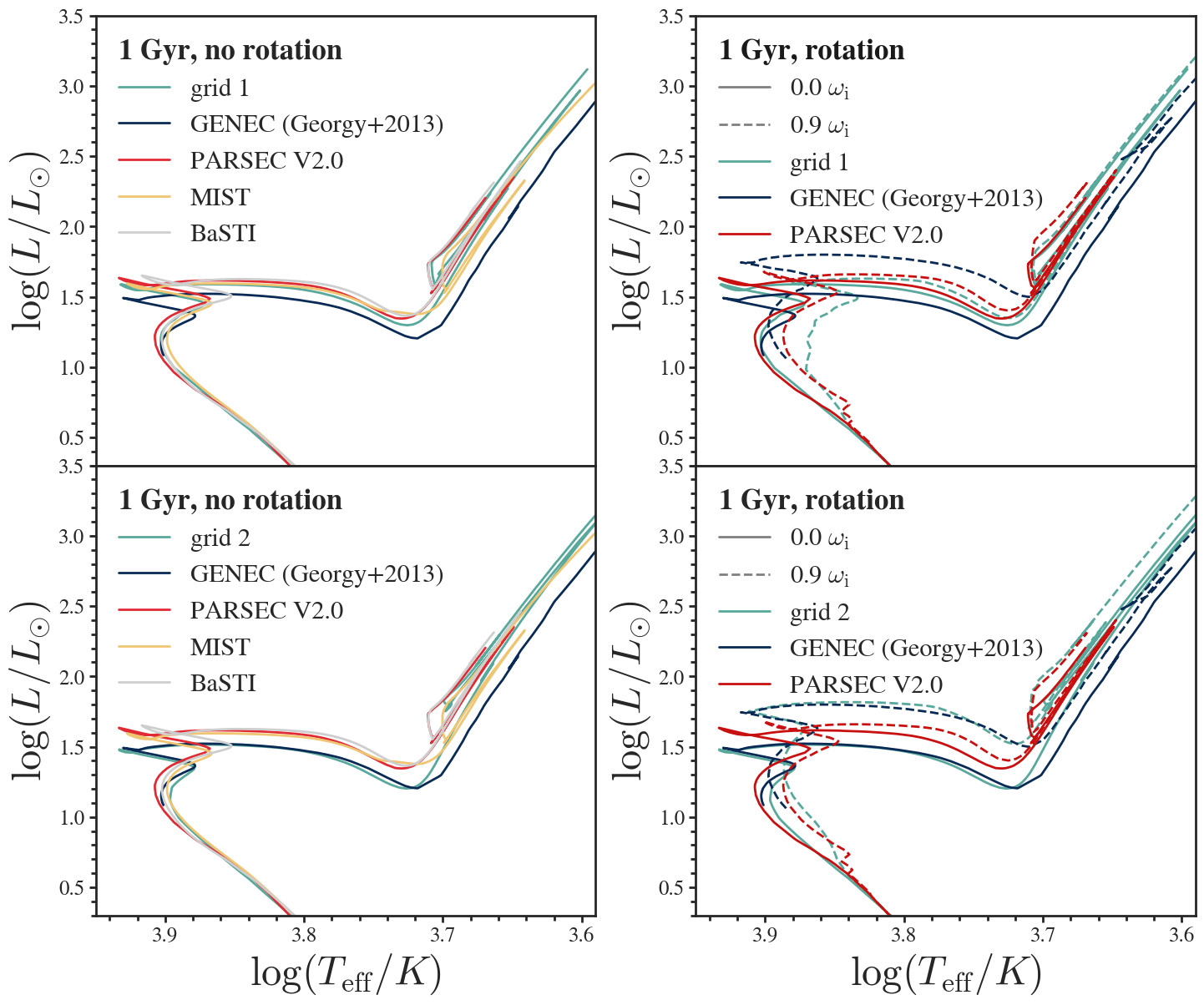}
    \caption{Left: Non-rotating 1 Gyr isochrones at $Z = 0.010$ from \texttt{grid 1} (top, green) and \texttt{grid 2} (bottom, green), compared with \texttt{GENEC} \citep{Georgy+2013} (blue), \texttt{PARSEC V2.0} \citep{Nguyen+2022} (red), \texttt{MIST} \citep{Choi+2016} (yellow), and \texttt{BaSTI} \citep{Hidalgo+2018} (grey). The fainter MSTO in \texttt{GENEC} reflects its smaller core overshooting. Note the close match between \texttt{grid 2} and \texttt{GENEC}, as intended. Right: 1 Gyr isochrones at $Z = 0.010$ for non-rotating (solid) and rotating ($\omega_{\rm{i}} = 0.9$, dashed) models from \texttt{grid 1} (top, green) and \texttt{grid 2} (bottom, green), compared with \texttt{GENEC} (blue) and \texttt{PARSEC V2.0} (red). The rotating isochrones appear progressively hotter and more luminous from \texttt{grid 1} to \texttt{PARSEC V2.0} to \texttt{GENEC}, reflecting the increasingly efficient rotational mixing adopted in each set of models.}
    \label{fig:iso}
\end{figure*}

In the left panel of Fig.~\ref{fig:iso}, we compare 1~Gyr non-rotating isochrones computed using models from \verb|grid 1| (top left) and \verb|grid 2| (bottom left) with non-rotating isochrones from the \texttt{GENEC} \citep[][]{Georgy+2013}, \texttt{PARSEC V2.0} \citep{Nguyen+2022}, \texttt{MIST} \citep{Choi+2016}, and \texttt{BaSTI} \citep{Hidalgo+2018} databases. The \verb|grid 1| non-rotating isochrone is consistent with those from the other databases. In contrast, the \texttt{GENEC} models exhibit a significantly fainter MSTO due to the smaller convective core overshooting adopted. Note the close agreement between the \verb|grid 2| and \texttt{GENEC} isochrones, as intended.

In the right panel of Fig.~\ref{fig:iso}, we show a comparison between non-rotating and rotating 1~Gyr isochrones computed using models from \verb|grid 1| (top) and \verb|grid 2| (bottom), alongside the corresponding rotating isochrones from \texttt{GENEC} \citep[][]{Georgy+2013} and \texttt{PARSEC V2.0} \citep[][]{Nguyen+2022}. Note that here we use the theoretical definitions of $L$ and $T_{\rm{eff}}$, not the observationally derived quantities $L_{\rm{MES}}$ and $T_{\rm{eff, MES}}$.

The rotating \texttt{GENEC} isochrone (and, by construction, that from \verb|grid 2|) is hotter and more luminous than the rotating \texttt{PARSEC V2.0} isochrone, which in turn is hotter and more luminous than that from \verb|grid 1|. This reflects the increasing strength of rotational mixing in the \texttt{GENEC}, \texttt{PARSEC V2.0}, and \verb|grid 1| models, respectively.
%%%%%%%%%%%%%%%%%%%%%%%%%%%%%%%%%%%%%%%%%%%%%%%%%%

% Don't change these lines
\bsp	% typesetting comment
\label{lastpage}
\end{document}